%% file: 11Be_PRC.tex
\documentclass[%
  superscriptaddress,
  showkeys,
  showpacs,
  reprint,
  nofootinbib,
  amsmath,amssymb,
  aps,
  prc,
  floatfix,
]{revtex4-1}

\usepackage{graphicx}
\usepackage{dcolumn}
\usepackage{bm}
\usepackage{multirow}
\usepackage{booktabs}

\begin{document}
\preprint{APS/123-QED}

\title{Reactions of a $^{10}$Be beam on proton and deuteron targets}

\author{K.T. Schmitt}
\altaffiliation[Current address ]{ORTEC/AMETEC Oak Ridge, Tennessee, TN37831-0895, USA}
\affiliation{Department of Physics and Astronomy, University of Tennessee, Knoxville, TN 37996, USA.}
\author{K.L Jones}
\email[Corresponding author: ]{kgrzywac@utk.edu}
\affiliation{Department of Physics and Astronomy, University of Tennessee, Knoxville, TN 37996, USA.}
\author{S. Ahn}
\affiliation{Department of Physics and Astronomy, University of Tennessee, Knoxville, TN 37996, USA.}
\affiliation{National Superconducting Cyclotron Laboratory and Department of Physics and Astronomy, Michigan State University, East Lansing, Michigan 48824, USA.}
\author{D.W. Bardayan}%
\affiliation{Physics Division, Oak Ridge National Laboratory, Oak Ridge, TN 37831, USA.}%
\affiliation{Department of Physics, Notre Dame University, Notre Dame, IN 46556, USA.}
\author{A. Bey}
\affiliation{Department of Physics and Astronomy, University of Tennessee, Knoxville, TN 37996, USA.}
\author{J.C. Blackmon}%
\affiliation{Department of Physics and Astronomy, Louisiana State University, Baton Rouge, LA 70803-4001 USA.}%
\author{S.M.~Brown}
\affiliation{Department of Physics, University of Surrey, Guildford, Surrey, GU2 7XH, UK.}
\author{K.Y.~Chae}%
\affiliation{Physics Division, Oak Ridge National Laboratory, Oak Ridge, TN 37831, USA.}%
\affiliation{Department of Physics, Sungkyunkwan University, Suwon 440-749, Korea.}
\affiliation{Department of Physics and Astronomy, University of Tennessee, Knoxville, TN 37996, USA.}
\author{K.A. Chipps}%
\affiliation{Physics Department, Colorado School of Mines, Golden, CO 80401, USA.}%
\affiliation{Department of Physics and Astronomy, University of Tennessee, Knoxville, TN 37996, USA.}
\author{J.A. Cizewski}
\affiliation{Department of Physics and Astronomy, Rutgers University, New Brunswick, NJ 08903, USA.}
\author{K. I. Hahn}
\affiliation{Ewha Womans University, 11-1 Daehyun-Dong Seodaemun-Gu Seoul 120-750, Korea.}
\author{J.J. Kolata}
\affiliation{Department of Physics, Notre Dame University, Notre Dame, IN 46556, USA.}
\author{R.L. Kozub}
\affiliation{Department of Physics, Tennessee Technological University, Cookeville, TN 38505, USA.}
\author{J.F.~Liang}
\affiliation{Physics Division, Oak Ridge National Laboratory, Oak Ridge, TN 37831, USA.}%
\author{C.~Matei}
\altaffiliation[Current address ] {Institute for Reference Materials and Measurements, Geel, Belgium.}
\affiliation{Oak Ridge Associated Universities, P.O. Box 2008, Oak Ridge, TN 37831, USA.}
\author{M.~Matos}
\affiliation{Department of Physics and Astronomy, Louisiana State University, Baton Rouge, LA 70803-4001 USA.}%
\author{D.~Matyas}
\affiliation{Department of Physics and Astronomy, Denison University, Granville, OH 43023, USA.}
\author{B.~Moazen}
\affiliation{Department of Physics and Astronomy, University of Tennessee, Knoxville, TN 37996, USA.}
\author{C.D.~Nesaraja}
\affiliation{Physics Division, Oak Ridge National Laboratory, Oak Ridge, TN 37831, USA.}%
\author{F.M.~Nunes}
\affiliation{National Superconducting Cyclotron Laboratory and Department of Physics and Astronomy, Michigan State University, East Lansing, Michigan 48824, USA.}
\author{P.D.~O'Malley}
\affiliation{Department of Physics and Astronomy, Rutgers University, New Brunswick, NJ 08903, USA.}
\affiliation{Physics Department, Colorado School of Mines, Golden, CO 80401, USA.}%
\author{S.D.~Pain}
\affiliation{Physics Division, Oak Ridge National Laboratory, Oak Ridge, TN 37831, USA.}%
\author{W.A.~Peters}
\altaffiliation[Current address]{ Joint Institute for Nuclear Physics and Applications, Oak Ridge, TN 37831.}
\affiliation{Department of Physics and Astronomy, Rutgers University, New Brunswick, NJ 08903, USA.}
\author{S.T. Pittman}
\affiliation{Department of Physics and Astronomy, University of Tennessee, Knoxville, TN 37996, USA.}
\affiliation{Department of Physics and Astronomy, Louisiana State University, Baton Rouge, LA 70803-4001 USA.}%
\author{A. Roberts}
\altaffiliation[Current address]{ Physics Division, Los Alamos National Laboratory, Los Alamos, NM 87545, USA.}
\affiliation{Department of Physics, Notre Dame University, Notre Dame, IN 46556, USA.}
\author{D. Shapira}
\affiliation{Physics Division, Oak Ridge National Laboratory, Oak Ridge, TN 37831, USA.}%
\author{J.F.~Shriner~Jr.}
\affiliation{Department of Physics, Tennessee Technological University, Cookeville, TN 38505, USA.}
\author{M.S.~Smith}
\affiliation{Physics Division, Oak Ridge National Laboratory, Oak Ridge, TN 37831, USA.}%
\author{I. Spassova}
\affiliation{Department of Physics and Astronomy, Rutgers University, New Brunswick, NJ 08903, USA.}
\author{D.W. Stracener}
\affiliation{Physics Division, Oak Ridge National Laboratory, Oak Ridge, TN 37831, USA.}%
\author{N.J. Upadhyay}
\affiliation{National Superconducting Cyclotron Laboratory and Department of Physics and Astronomy, Michigan State University, East Lansing, Michigan 48824, USA.}
\author{A.N. Villano}
\altaffiliation[Current address]{ School of Physics and Astronomy, University of Minnesota, Minneapolis, MN 55455, USA.}
\affiliation{Department of Physics, Western Michigan University, Kalamazoo, MI 49008, USA.}
\author{G.L. Wilson}
\altaffiliation[ Current address]{ Department of Physics, University of York, Heslington, York, UK, YO105DD.} %
\affiliation{Department of Physics, University of Surrey, Guildford, Surrey, GU2 7XH, UK.}%

\date{\today}


\begin{abstract}
The extraction of detailed nuclear structure information from transfer reactions requires reliable, well-normalized data as well as optical potentials and a theoretical framework demonstrated to work well in the relevant mass and beam energy ranges.  It is rare that the theoretical ingredients can be tested well for exotic nuclei owing to the paucity of data.  The halo nucleus $^{11}$Be has been examined through the $^{10}$Be(d,p) reaction in inverse kinematics at equivalent deuteron energies of $12,15,18,~$and$~21.4$~MeV.  Elastic scattering of $^{10}$Be on protons was used to select optical potentials for the analysis of the transfer data.  Additionally, data from the elastic and inelastic scattering of $^{10}$Be on deuterons was used to fit optical potentials at the four measured energies. Transfers to the two bound states and the first resonance in $^{11}$Be were analyzed using the Finite Range ADiabatic Wave Approximation (FR-ADWA).  Consistent values of the spectroscopic factor of both the ground and first excited states were extracted from the four measurements, with average values of 0.71(5) and 0.62(4) respectively.  The calculations for transfer to the first resonance were found to be sensitive to the size of the energy bin used and therefore could not be used to extract a spectroscopic factor.
  
 \begin{description}
 \item[PACS numbers]
\pacs{25.60.Je, 25.60.Bx, 25.45.De, 25.45.Hi}
\keywords{Direct reactions, transfer reactions, elastic scattering, inelastic scattering, radioactive ion beams}
 \end{description}
\end{abstract}

\maketitle


\section{\label{sec:intro}Introduction:\protect\\ }

Rare isotope beam facilities have made it possible to study nuclei at the limits of stability and beyond in the light neutron-rich region of the nuclear chart.  The imbalances of proton number and neutron number and decreased particle separation energies that occur in such regions lead to several exotic phenomena.  Neutron halos are a notable example of these effects, wherein weakly bound valence neutrons reside in diffuse, spatially extended distributions.

Since the discovery of the ground-state halo of $^{11}$Be \cite{Tan88}, there has been a lot of interest in understanding its structure.  Unsurprisingly, the structure of $^{11}$Be has been studied via many different methods including $\beta$ decay, e.g.  \cite{Hir05, Mad09, Sar04},  neutron knockout \cite{Aum00}, nuclear and Coulomb breakup \cite{Pal03,Fuk04,Lim07}, and one- and two-neutron transfer reactions \cite{Pul62,Aut70,Ajz72,Ajz78,Zwi79,Liu90, For99, Win01,Hai09}.  There has also been  intense interest on the theoretical side with various studies of the structure and reactions with $^{11}$Be, such as \cite{Ots93, Esb95, Suz95, Vin95, Vin96, Nun96, Des97, Bha97, For05, Del09, Del13}, some of which are summarized by Winfield et al \cite{Win01}.  In some of these studies the ground state spectroscopic factor was calculated, resulting in values ranging from 0.55 \cite{Ots93} to 0.93 \cite{Vin95}.   The situation regarding the spectroscopic factors of the bound states was similarly complicated on the experimental side as summarized by Fortune and Sherr \cite{For12}.

Values of the spectroscopic factors  of states in $^{11}$Be have been extracted from the two normal kinematics $^{10}$Be(d,p)$^{11}$Be measurements \cite{Aut70, Zwi79},  from widths of resonances following the  $^{9}$Be(t,p)$^{11}$Be reaction \cite{Liu90}, and from beta-decay  \cite{Hir05}.  The two (d,p) results agree for the ground state spectroscopic factor, but not for the first excited state (see table \ref{spec}).  Extracted spectroscopic factors from transfer reactions are sensitive to a number of factors, including the optical and bound state potentials, as well as the reaction model used.  Both original sets of data were analyzed using either the standard Distorted Wave Born Approximation (DWBA) method or a modified version that allows two-step transfer, Coupled Channels Born Approximation (CCBA)  \footnote {Zwieglinski et al performed CCBA calculations and found that the effect on spectroscopic factors was small(10\% for excited states, 20\% for ground state).}.  These are not ideal methods of reaction analysis especially when there are two weakly-bound nuclei involved, as DWBA does not explicitly include particle break-up.  Subsequently, the data were reanalyzed by Keeley, using the continuum discretized coupled channels (CDCC) procedure to model deuteron breakup \cite{kee04}.  The discrepancy in the excited state spectroscopic factor persists in this analysis, suggesting that either there is a problem with one or both sets of data, or that there are effects that are not being included correctly in the calculations.
\begin {table*}
\caption{\label{spec}
Compiled J$^{\pi}$ assignments and spectroscopic factors of states below E$_{x}$ = 3.5~MeV in $^{11}$Be from experiment.  Selected spectroscopic factors from theory are included for comparison. }

\begin{ruledtabular}
\begin{tabular}{c c c c c} 
& &   \multicolumn{2}{c} {Spectroscopic Factor (S) from Experiment}&{Spectroscopic Factor (S)} \\
& & 									&				& {from Theory} \\
\cline {3-5}
& & & Nuclear Breakup or & \\
{E$_x$} & {J$^{\pi}$}& Transfer&Coulomb Dissociation & \\
\hline
   0	          & 1/2$^+$ 		& 0.73$\pm 0.06$, 0.77 \cite{Aut70,Zwi79}	& 0.87$\pm 0.13$, 0.61$\pm 0.05$, 	&	0.964, 0.78 \cite{Vin95,Nun96} \\
   		 &				&									&0.72$\pm0.04$, 0.46$\pm0.15$ \cite{Aum00,Pal03,Fuk04,Lim07}							&			\\
   0.320	 & 1/2$^-$ 		& 0.63$\pm 0.15$, 0.9 \cite{Aut70,Zwi79}		& -																			&	0.746, 0.87 \cite{Vin95, Nun96} \\
   1.778    	 & 5/2$^+$ 		& 0.50, 0.58(8) \cite{Zwi79,For11}			& -																			&	0.896 \cite{Vin95}	 \\
   2.69	 & 3/2$^-$ 		& $\approx0.12$ \cite{For11}				& -																			&	0.168 \cite{Coh71}	 	\\
   3.41	 & [3/2$^-$, 3/2$^+$]	& $\approx0.05$ \cite{For11}				& -																			&		- \\
\end{tabular}
\end{ruledtabular}
\end{table*}

On the experimental side, it is essential that the method of normalizing the data to obtain absolute cross sections is robust.  The target used for the E$_d$~=~12~MeV data \cite{Aut70} was not well characterized, making an absolute normalization exceedingly difficult.  The authors normalized the data using elastic scattering data compared to DWBA calculations.  This was necessary as the measurement was made at an energy well above the Rutherford regime, and resulted in a model dependency in the absolute cross-sections.  

Here we discuss measurements using a $^{10}$Be beam on targets of deuterons at four beam energies; E$_{10}$~=~60~MeV, 75~MeV, 90~MeV, and 107~MeV equivalent to E$_d$~=12~MeV, 15~MeV, 18~MeV, and 21.4~MeV, and targets of protons at the same $^{10}$Be beam energies, i.e. at equivalent proton energies of E$_p$~=6~MeV, 7.5~MeV, 9~MeV, 10.7~MeV.  The data from these measurements are tabulated in the Supplemental Material \cite{sup13}. Some of the results from the (d,p) and (d,d) reactions in inverse kinematics have been reported in \cite{Sch12}.  Those data were subsequently reanalyzed using an exact Faddeev-type framework by Deltuva \cite{Del13} where qualitatively good agreement with the data was found; however spectroscopic factors could not be extracted owing to the non-scalable nature of the theory. The current paper presents further details of the original analysis, as well as data for the inelastic scattering and reactions on protons.  Additionally,  data from the transfer to the first resonance in $^{11}$Be will be presented.  

The data are analyzed within the finite range ADiabatic Wave Approximation (ADWA) formalism in a consistent manner. The spectroscopic factors of the bound and first resonant states extracted from each measurement are compared to each other, to previous measurements, and to theoretical predictions.  In this way the uncertainties arising from experimental factors, from choices of scattering potentials, or from the theoretical framework itself can be isolated.  This analysis is a first step to a complete interpretation of the data.  

Unlike the Faddeev-type methods, the ADWA formalism is an approximation and does not include higher-order effects such as core excitation.  Such a complete set of data, as presented here, provides an excellent test bed for exploring the validity and limitations of different theoretical frameworks.
\begin{figure*}
\includegraphics[width=5cm]{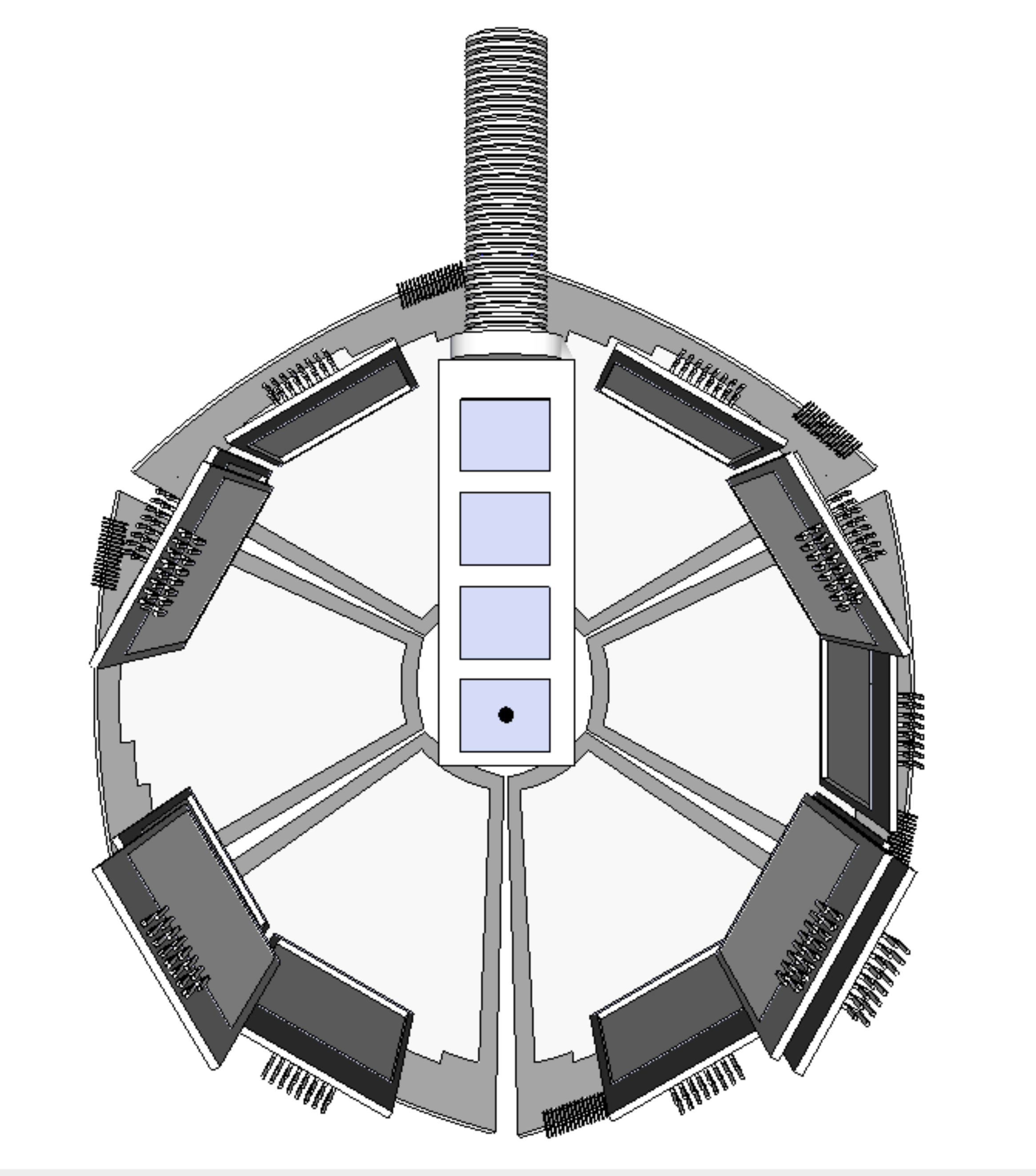}
\includegraphics[width=8cm]{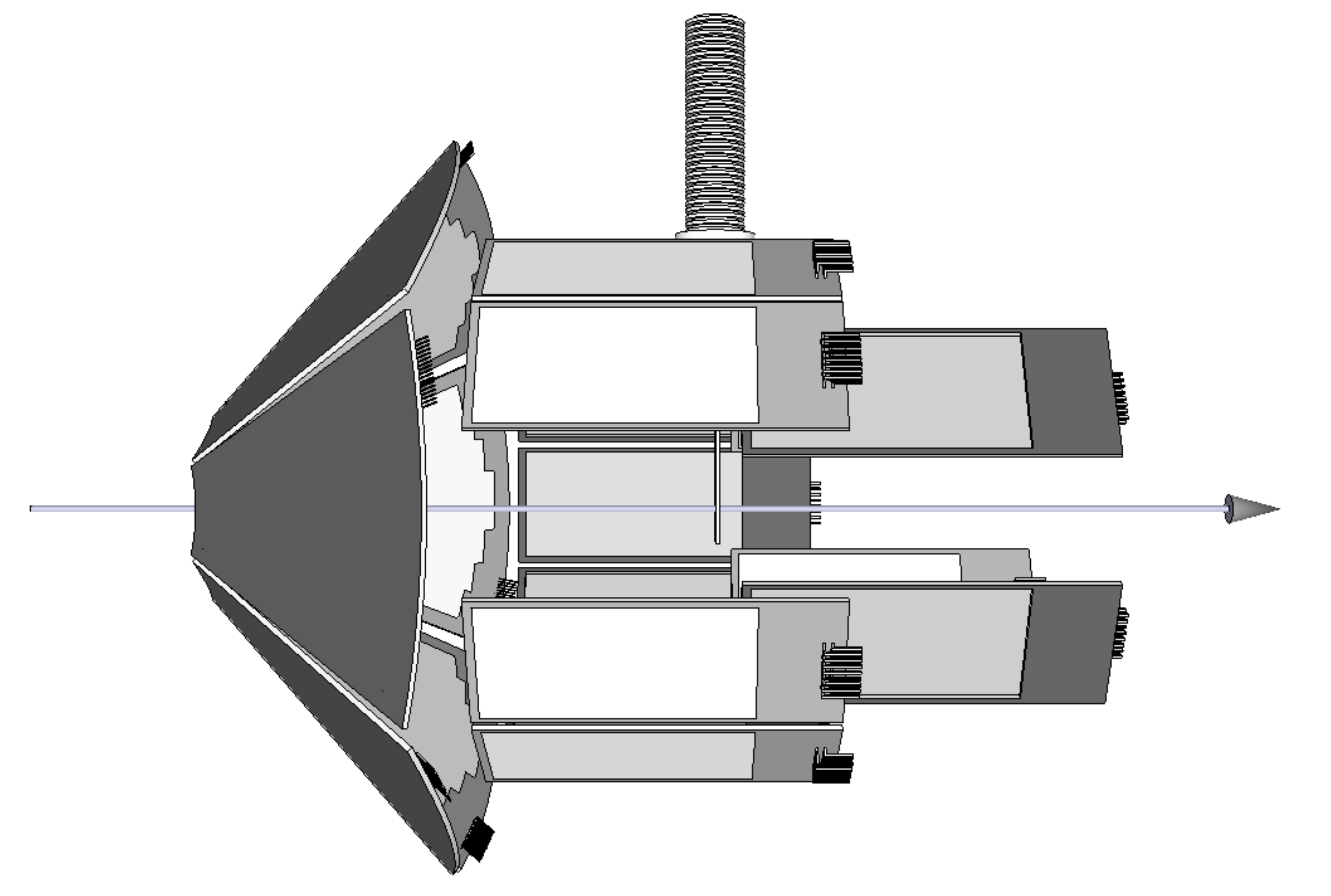}
\caption{\label{setupfig}Front and side views of the silicon detector setup used for light ion detection in the $^{10}$Be + d measurements (not to scale).  Beam direction is indicated by an arrow in the side view.  SIDAR \cite{Bar99} was mounted in a lampshade configuration at backward laboratory angles ($138^{\circ}$  $\le$ $\theta_{lab}$ $\le$ 165$^{\circ}$), and ORRUBA covered an angular range of $45^{\circ}$  $\le$ $\theta_{lab}$ $\le$ 135$^{\circ}$ \cite{Pai07}.  SuperORRUBA \cite{Bar13} detectors were used at forward laboratory angles for the proton scattering measurements. \protect\\  }
\label{setup}
\end{figure*}
\section{\label{sec:level1}Experimental Setup:\protect\\ }
Data were collected at the Holifield Radioactive Ion Beam Facility (HRIBF) \cite{Bee11} at Oak Ridge National Laboratory.  Long-lived $^{10}$Be ions were accelerated from a cesium sputter source using the 25 MV tandem accelerator.  The source was prepared from a solution of $^{10}$Be in hydrochloric acid.  Contamination from residual $^{10}$B material in the ion source was removed by fully stripping the beam ions in the accelerator and tuning the final analyzing magnet for Z = 4.  The negligible level of contamination was confirmed by the non-observation of kinematic lines from $^2$H($^{10}$B,p)$^{11}$B reactions at E$_d=21.4$ MeV, which were clearly evident at backward angles in the laboratory frame with a pure $^{10}$B beam.  For subsequent runs, beam contamination was ruled out by identifying beam particles using an ionization chamber.  By using a batch mode source - that is one that produces a long-lived radioactive ion beam in a similar way to a stable ion beam - and a tandem accelerator, high quality, high statistics measurements could be made.  In particular, the beam accelerated in a tandem accelerator has low emittance, which allows precise reconstruction of the kinematics of the reaction.

The CD$_2$ targets were prepared from deuterated polyethylene powder with areal densities of 94, 143, 162, and 185~$\mu$g/cm$^2$.  The CH$_2$ target was prepared from commercially available polyethylene sheets, stretched to achieve an areal density of 226~$\mu$g/cm$^2$.  The thickness of each target was measured using the energy loss of 5.805 MeV $\alpha$-particles from the decay of $^{244}$Cm.

The experiment was performed in three parts.  The initial part was performed at a beam energy of 107~MeV using a deuterated target (E$_d=21.4$~MeV).  In the second part, runs were performed at 60, 75, and 90 MeV (E$_d$=12, 15, and 18~MeV).  The final runs were performed using the same four beam energies and CH$_2$ targets (E$_p=6$, 7.5, 9, and 10.7~MeV). 

A new Dual Micro-Channel Plate (DMCP) detector was employed for heavy recoil detection during the first part.  The DMCP utilizes two conventional MCP detectors \cite{Shapira}, viewing opposing sides of a single carbon or metallized foil, thereby increasing the measurement efficiency, particularly for low-Z beams where the single MCP efficiency can be well below 50\%.  A benefit of this technique is that the counting efficiency of the DMCP can be verified easily by comparing singles and coincidence rates. The DMCP 
was replaced in the later runs by a new fast ionization chamber, similar in design to the Tilted Electrode Gas Ionization Chamber \cite{kim05}.  

For the data taken with deuterated targets, the angles and energies of light ion ejectiles were measured using the Silicon Detector Array (SIDAR \cite{Bar99}) and the first full implementation of the Oak Ridge Rutgers Barrel Array (ORRUBA \cite{Pai07}) as shown in Fig. \ref{setup}.  ORRUBA detectors (1000~$\mu$m thick) were mounted at a radius of 87~mm from the beam axis at $\theta_{lab}>95^{\circ}$ and at 76 mm at more forward angles.  The SIDAR array was mounted in lampshade configuration at backward angles to cover angles between $135^{\circ}$ and $165^{\circ}$.  This combination yielded nearly continuous solid angular coverage, except for a small gap due to shadowing from the target ladder.  SuperORRUBA detectors \cite{Bar13}, which are segmented for 1.2~mm position sensitivity, were used at angles forward of $90^{\circ}$ for the proton scattering measurements.  

For the initial measurement at 107~MeV, $\Delta$E-E telescopes of ORRUBA detectors were employed at forward angles for particle identification and an additional array of silicon detectors with annular strips was used to identify beam-like recoils at far forward angles ($1^{\circ}<\theta<8^{\circ}$).  Carbon atoms scattered from the target were stopped in a 16~mg/cm$^2$ polymer film placed in front of one of the forward angle ORRUBA detectors.  

An angular resolution of better than 2$^{\circ}$ in polar angle was achieved throughout the angular coverage of the detectors.  Energies were calibrated using a 5.805 MeV $\alpha$ source.  Pulser data were used to determine the zero energy channel for the electronics.

To normalize the data, absolute elastic scattering cross sections were measured for each target species and beam energy using ORRUBA or Super ORRUBA detectors.  In subsequent measurements, the reaction channels of interest were recorded simultaneously with elastic scattering so that absolute cross sections could be determined with reference to the absolute elastic scattering cross section measurements.  

The absolute elastic scattering cross section measurements were performed with reduced beam intensities to allow for reliable beam counting.  Fresh targets were used to minimize the effects of target degradation.  A sieve-type attenuator was placed in front of the ionization chamber for the deuteron scattering runs to make it possible to measure beam rates up to $10^6$ particles per second.  The transmission through the attenuator was measured by passing the beam through a thin silicon detector with a beam rate of $\approx 10^3$ pps before reaching the attenuator.  It was found that one if every $7.4 \pm 0.1$ incident particles was transmitted.  Transmission measurements were taken at several places on the attenuator and found to be accurate to within 2\%.   Systematic uncertainties for the normalization of the deuteron scattering data are shown in Table \ref{tabuncertd}. \\

A double-layered attenuator was employed for the proton scattering data taken at $E_p=7.5$ MeV.  The rate of attenuation in that case was found to be $39.1 \pm 2.7$ by comparing data taken with and without the attenuator.  No attenuator was used for the proton scattering normalization runs at other energies.  This resulted in an increase in systematic uncertainty for the data taken at $E_p=7.5$ MeV relative to the other proton scattering data.  Systematic uncertainties for the normalization of the proton scattering data are shown in Table \ref{tabuncertp}. 

\begin {table}
\caption{\label{tabuncertd}
Breakdown of systematic uncertainties in the differential cross sections from the deuteron-induced reactions.  Uncertainties are quoted as percentages.
}
\begin{ruledtabular}
\begin{tabular}{lcccc} 
&\multicolumn{4}{c}{E$_d$ (MeV)}\\
\cline{2-5}
\textrm{Source of uncertainty}&12&15&18&21.4\\
\colrule
   Solid angle coverage\footnotemark[1]		&	3	&	3	&	3	&	3	\\
   Sieve rate\footnotemark[2]				&	1	&	1	&	1	&	0	\\
   Beam counting measurement\footnotemark[2]	&	5	&	5	&	5	&	5	\\
   Target thickness\footnotemark[3]			&	5	&	5	&	5	&	5	\\ 
   Normalization run statistics     	&	6	&	6	&	6	&	4	\\
   Normalization gating	     	&    5	&	5	&	5	&	5	\\
\colrule
   Total uncertainty &	11	&	10	&	10	&	10
\end{tabular}
\footnotetext[1]{This uncertainty is common to the data taken at each energy.}
\footnotetext[2]{This uncertainty is common to the data taken at 12, 15, and 18 MeV.}
\footnotetext[3]{This uncertainty is common to the data taken at 12 and 18 MeV.}
\end{ruledtabular}
\end{table}

\begin {table}
\caption{\label{tabuncertp}
Breakdown of systematic uncertainties in the differential cross-sections from the proton-induced reactions.  Uncertainties are quoted as percentages.
}
\begin{ruledtabular}
\begin{tabular}{lcccc} 
&\multicolumn{4}{c}{E$_p$ (MeV)}\\
\cline{2-5}
\textrm{Source of uncertainty}&6&7.5&9&10.7\\
\colrule
   Solid angle coverage\footnotemark[1]		&	3	&	3	&	3	&	3	\\
   Sieve rate							&	0	&	7	&	0	&	0	\\
   Beam counting measurement\footnotemark[2]	&	5	&	5	&	5	&	5	\\
   Target thickness\footnotemark[3]			&	5	&	5	&	5	&	5	\\ 
   Normalization run statistics     	&	3	&	5	&	3	&	2	\\
\colrule
   Total uncertainty &	8	&	12	&	8	&	8
\end{tabular}
\footnotetext[1]{This uncertainty is common to the data taken at each energy.}
\footnotetext[2]{This uncertainty is common to the data taken at 6, 7.5, and 9 MeV.}
\footnotetext[3]{This uncertainty is common to the data taken at 6 and 9 MeV.}
\end{ruledtabular}
\end{table}

\section{ELASTIC AND INELASTIC SCATTERING}


Optical potentials are an important component of the analysis of transfer data. Since the ADiabatic Wave Approximation (ADWA) transfer analysis used here is based on a three-body approach to the reaction \cite{Ngu10}, nucleonic optical potentials are necessary ingredients.  Global nucleonic optical potentials are developed by fitting to a large collection of elastic scattering data, whereas non-global potentials are fitted to data for a particular nucleus.  In both cases the potentials are typically fitted to data from stable nuclei. Uncertainties associated with these potentials, especially when used for unstable nuclei, can be minimized by either creating a potential for the specific nucleus of interest by fitting the relevant elastic scattering cross sections, or by selecting global optical potentials by their ability to reproduce the measured elastic scattering data.     

\subsection{Proton elastic scattering}
\begin{figure}
\centering
\includegraphics[width=8.5cm]{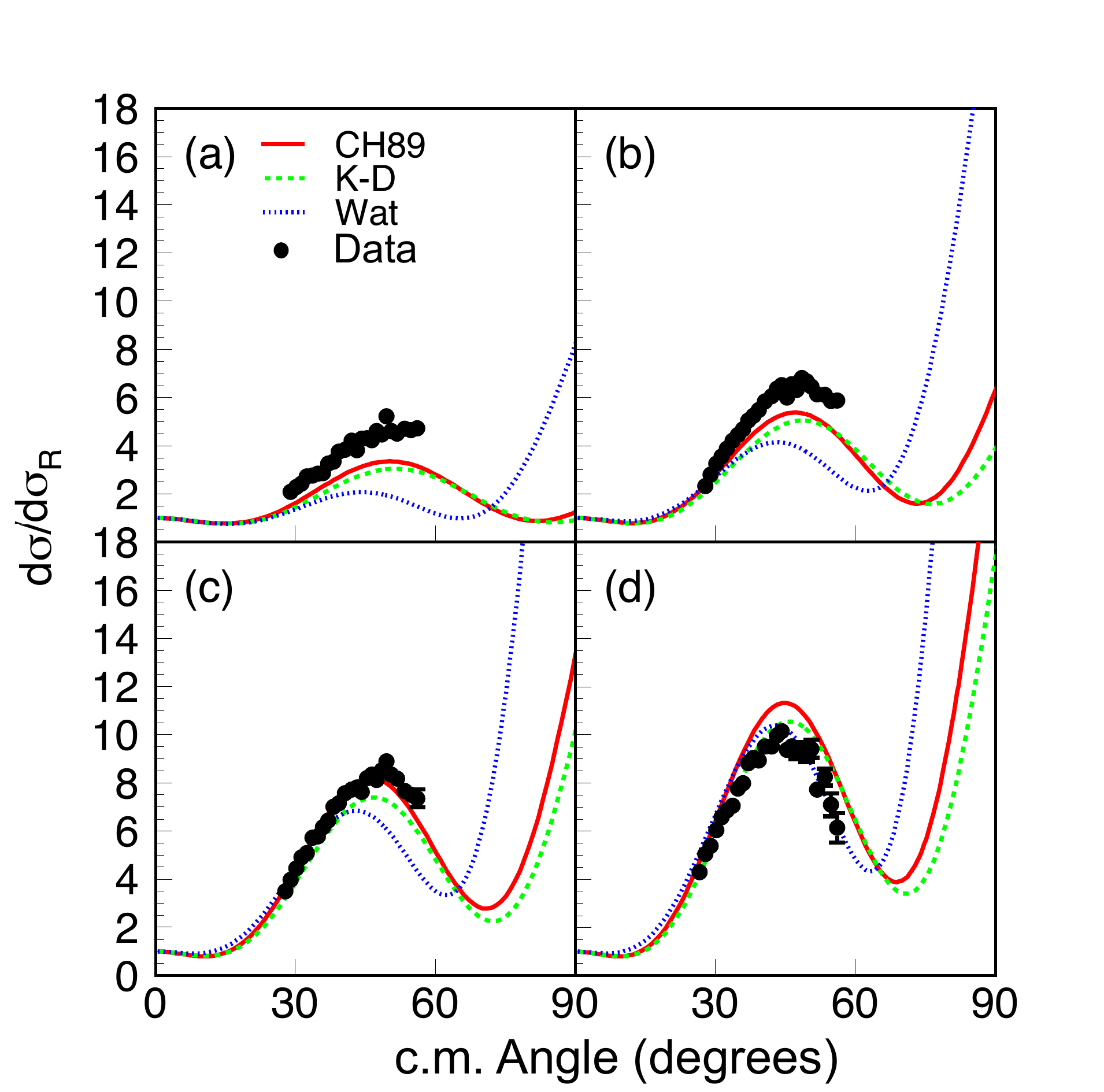}
\caption{(color online).  Differential cross-sections, shown as a ratio to Rutherford scattering, for proton elastic scattering at equivalent proton energies of E$_p$ =(a) 6, (b) 7.5, (c) 9, and (d) 10.7~MeV.}\label{pelasticplot}
\end{figure}

Elastic scattering cross sections were measured for protons with $^{10}$Be beams at energies identical to the transfer measurements to provide constraints on optical potentials both for the current study and for future reaction studies with light neutron-rich nuclei.   Angular distributions, Fig.~\ref{pelasticplot}, were determined by dividing the data into angular bins, and fitting the resulting one-dimensional histograms with Gaussian curves (more details of the data analysis will be given in section \ref{transfer}).  Background from fusion-evaporation on $^{12}$C was taken into account.  The uncertainties shown include contributions from statistics and ambiguities in curve fitting.   

Optical model calculations for the proton elastic scattering were performed with three potentials.  The proton-nucleus potentials employed were those of Varner et al (CH89) \cite{Var91}, Koning and Delaroche (K-D) \cite{Kon03}, and Watson et al (Wat) \cite{Wat69}.  No single optical potential from those studied can reproduce the proton elastic scattering data over this range of beam energies.  At the lower beam energies, Fig.~\ref{pelasticplot}a and b, CH89 and K-D potentials come closest to explaining the data; however, the calculations underestimate the intensity by up to 20\%.  At E$_p$~=~9~MeV, displayed in Fig.~\ref{pelasticplot}c, CH89 and K-D both reproduce the data well.  However, at the highest energy, Fig.~\ref{pelasticplot}d, CH89 overestimates the intensity of the proton elastic scattering, but Wat and K-D perform well. Given the divergence of the various models beyond 60 degrees, it would be desirable to measure the elastic channel in this angular range to help further constrain the nucleon optical potential.

\subsection{Deuteron elastic and inelastic scattering}
The deuteron target data included reactions beyond simple elastic deuteron scattering. Elastically and inelastically scattered deuterons, as well as the protons from (d,p), and elastically scattered target contaminant  protons are evident.   The data were analyzed in a similar method to that described above for the proton elastic scattering.  Deuteron elastic scattering cross sections were calculated with the potentials of Fitz \cite{Fit67}, Satchler (Sat) \cite{Sat66}, and Perey and Perey (P-P) \cite{Per63}.   

\begin{figure}
\centering
\includegraphics[width=8.5cm]{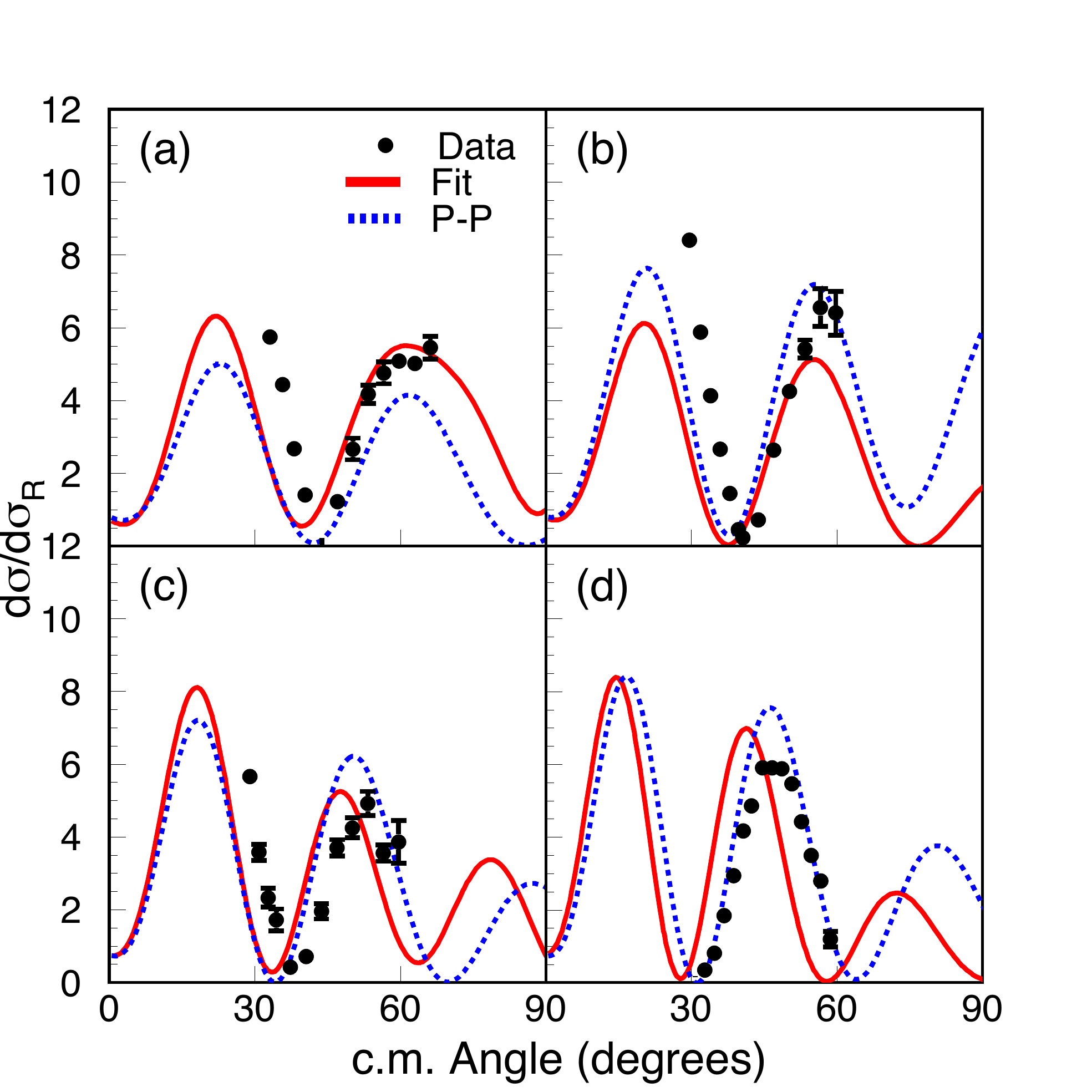}
\caption{(color online).  Differential cross-sections, shown as a ratio to Rutherford scattering, for deuteron elastic scattering at equivalent deuteron energies of E$_d$ = (a)~12, (b)~15, (c)~18, and (d)~21.4~MeV.}\label{delasticplot}
\end{figure}

\begin{figure}
\centering
\includegraphics[width=9.cm]{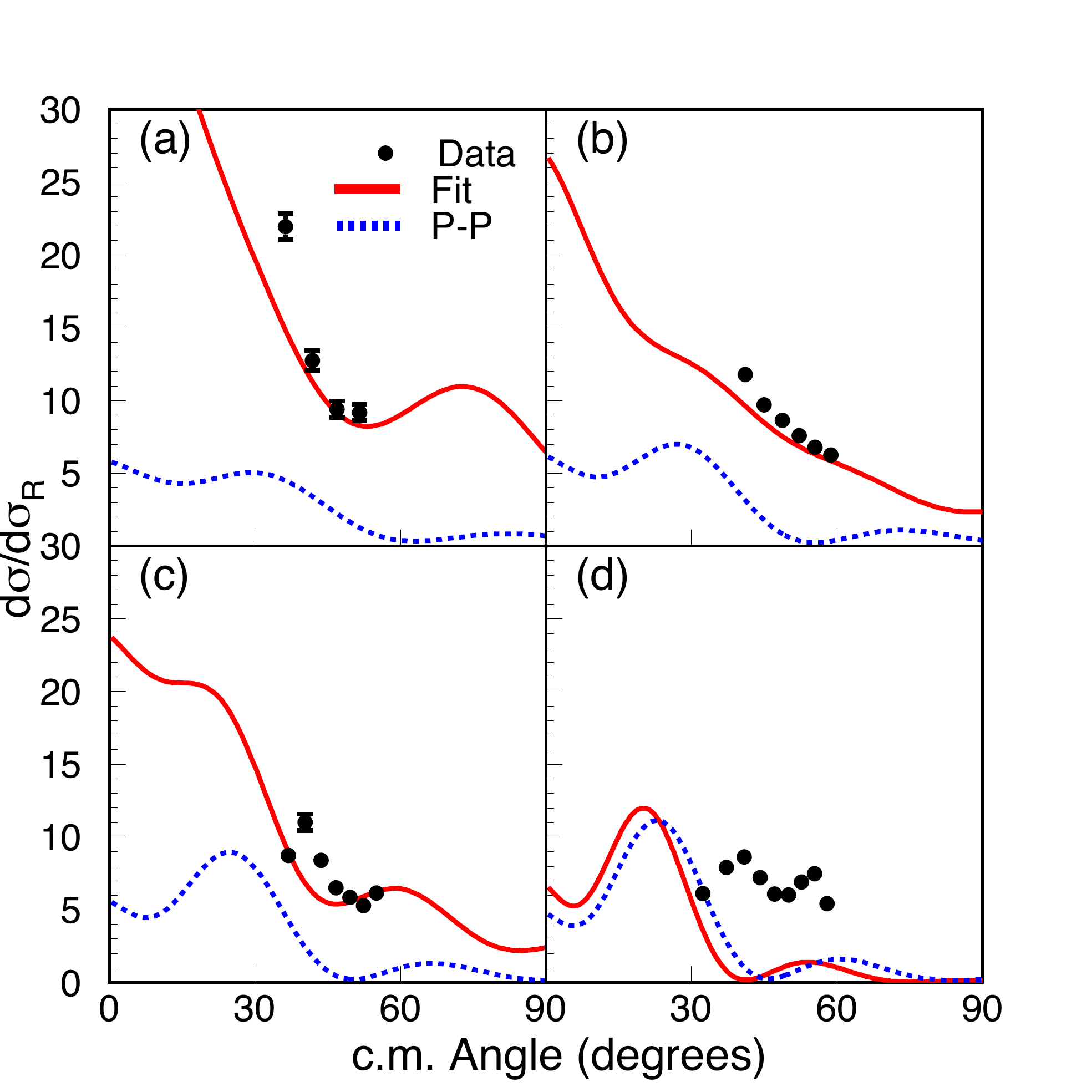}
\caption{(color online).  Differential cross-sections, shown as a ratio to Rutherford scattering, for deuteron inelastic scattering at equivalent deuteron energies of E$_d$ = (a)~12, (b)~15, (c)~18, and (d)~21.4~MeV.}\label{inelasticplot}
\end{figure}

In reference \cite{Sch12}, the deuteron elastic scattering data were first presented and analyzed with global deuteron optical potentials. It was observed that the first peak of the elastic distribution was generally consistent between the different optical potentials within the range of data available, while the second peak  was found to be more discriminating between the various global potentials available.   It was also shown in reference \cite{Sch12} that none of the optical potentials were able to reproduce the height of the second peak at E$_d$ = 15~MeV.  All in all, the results shown in Fig.~4 of reference \cite{Sch12} demonstrated the shortcoming of the global optical model for these reactions.  Here we go beyond the analysis of \cite{Sch12} and fit a potential to both the elastic and inelastic deuteron scattering data, shown in Figs.~\ref{delasticplot} and \ref{inelasticplot} respectively.

Although a full three-body approach including excitation of $^{10}$Be would be the preferred framework, it is beyond the scope of this work. Here we analyze the deuteron elastic and inelastic channels consistently within a coupled channel framework. We fix the quadrupole coupling to the $2^+$ $^{10}$Be state using a nuclear deformation length from reference \cite{Nun96}. This deformation length $\delta_2=1.84$ fm is consistent with the measurement of Iwasaki et al. \cite{Iwa00}. Including  deformation on top of one of the deuteron global optical potentials, and then performing a coupled channel calculation worsens the elastic scattering angular distributions and yields inelastic scattering angular distributions with no resemblance to the data. It is well known that in a coupled channel framework it is the bare interaction that needs to be included and in this sense, one needs to refit the optical potential when couplings are included. A fit to the elastic scattering distribution was performed through $\chi^2$ minimization for each energy, using the Perey-Perey global deuteron potential as a starting point and keeping the Coulomb radius fixed.  The resulting parameters are further optimized to jointly fit the inelastic distribution. Calculations were performed using the reaction code {\sc fresco}  and fitting code {\sc sfresco} \cite{Tho88}.  The resulting optical potential parameters are presented in Table~\ref{table-op} along with the corresponding $\chi^2$. 

Compared to the original potential, the depth of the real part of the interaction is significantly reduced for all of these cases, and the radius is increased, common features of effective potentials involving loosely bound systems. For all but the highest energy, the depth of the imaginary part was decreased, thereby compensating for the explicit inclusion of the inelastic channel. It was particularly difficult to obtain a good quality description for the $E_d=21$ MeV case, demonstrating that other channels, beyond the inelastic $2^+$ state, need to be considered. At this high energy, the data showed preference for a volume term for the imaginary part of the interaction, as opposed to all other energies where only a surface term provided a good description. 

The data, along with coupled channel angular distributions using the new fit, are presented in Fig.~\ref{delasticplot}, for the deuteron elastic scattering cross section relative to the Rutherford cross section, and in Fig.~\ref{inelasticplot} for the inelastic scattering cross section. The calculations using the original P-P global optical potential are shown for comparison.  The agreement with the elastic data is of similar quality for the new fitted potential compared to P-P; however, the fit provides much better agreement with the data in the case of the inelastic scattering, except at the highest energy, as described above.

\begin{table}
\center
\begin{tabular}{lcccccccccrr}\hline\hline
$E_d$ & $V$ & $r_r$ & $a_r$ & $W$ & $W_d$ & $R_i$ & $a_i$ & $\chi^2_{el}$  & $\chi^2_{inel}$ \\
 (MeV)&(MeV)&(fm)&(fm) &(MeV)& (MeV) &  (fm) &			 &  				& \\\hline
12 & 68.8 & 1.4 &0.78 & 0. & 9.3   & 1.0 &0.5 & 16.5 & 7.3  \\
15 & 68.8 & 1.7 &0.52 & 0.  & 9.1& 1.5 &0.85 & 25.9 & 1.2   \\
18 & 60.0   & 1.6 &0.63 & 0. & 7.4 & 1.6 & 0.68 & 2.7 & 10.8 \\
21 & 66.4  & 1.1 &0.90 & 17.8 & 0.     &1.7 &0.9 & 31.6 & 42.6   \\
\hline\hline
\end{tabular}
\caption{Optical potentials fit to d+ $^{10}$Be elastic and inelastic scattering data.}
\label{table-op}
\end{table}


\section{THE NEUTRON TRANSFER REACTION $^{10}\textrm{Be(d,p})$}\label{transfer}
\begin{figure}
\centering
\includegraphics[width=8.cm]{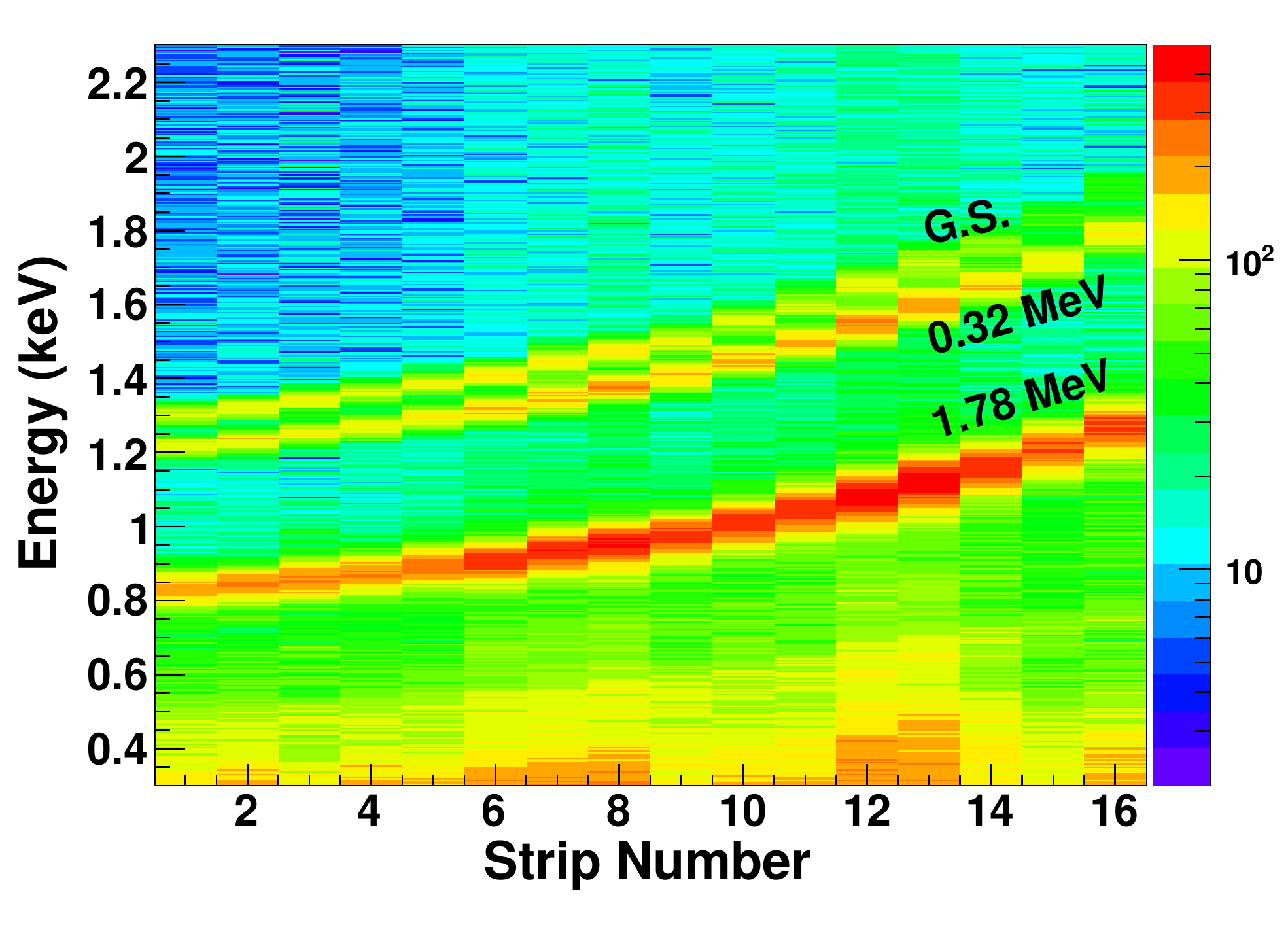}
\caption{(color online).  Proton energy in keV in the laboratory frame as a function of strip number for SIDAR at backward angles at an equivalent deuteron energy, E$_d$ = 21.4~MeV. }\label{2dtransfer}
\end{figure}

Angular distributions of protons emerging from (d,p) reactions with low orbital angular momentum transfer are typically peaked at small center-of-mass angles.  For this reason most of the data for the neutron-transfer channel comes from the SIDAR array, which was placed at backward angles in the laboratory frame.  Fig.~\ref{2dtransfer} shows the energies of protons measured in SIDAR at E$_d=21.4$~MeV as a function of angle, represented by the strip number, where strip 1 covers the largest laboratory angles.  Contours of constant Q-value, indicating the population of a discrete state, have been labelled.  

\begin{figure}
\centering
\includegraphics[width=8.cm]{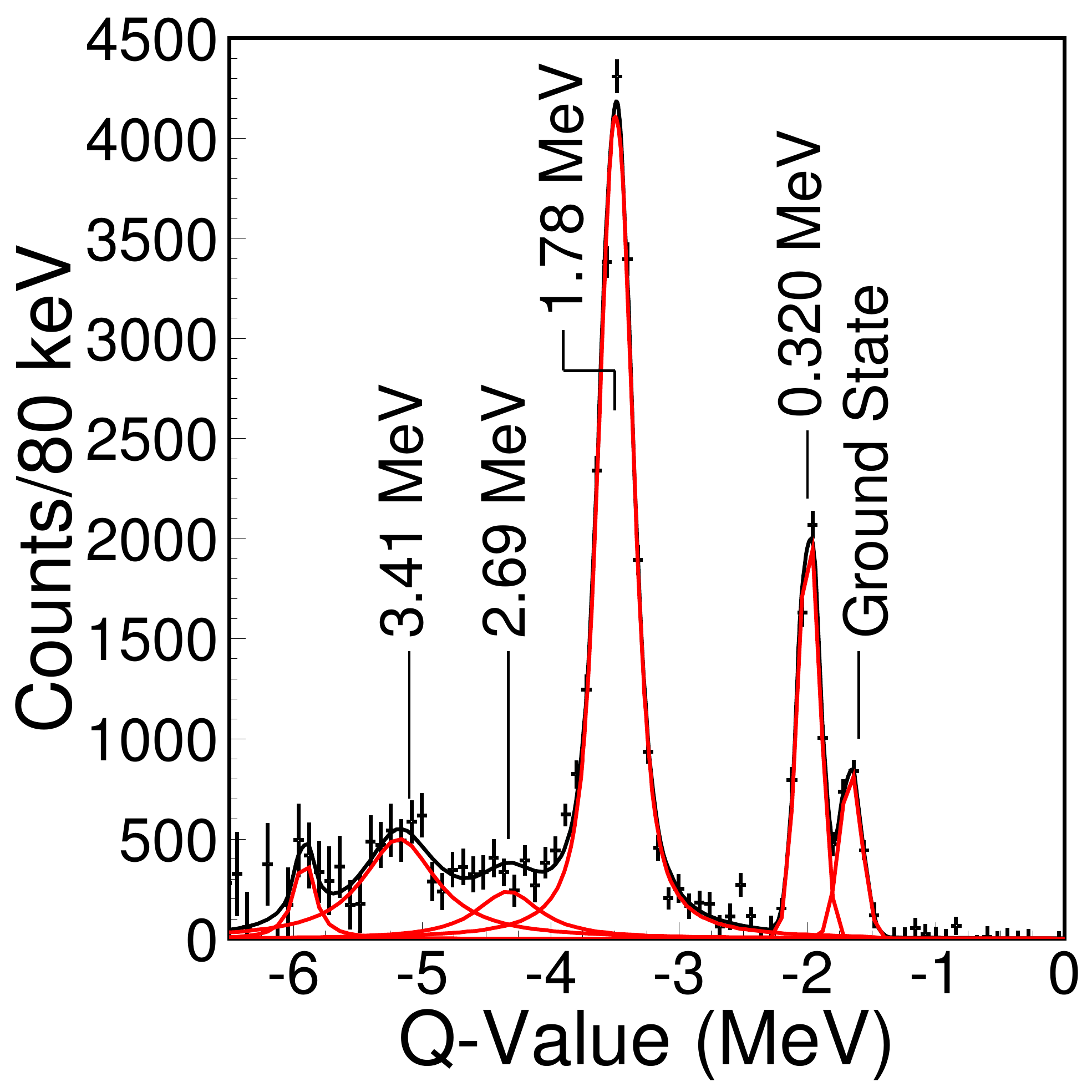}
\caption{(color online).  Q-value spectrum for $^{10}$Be(d,p)$^{11}$Be in inverse kinematics at 140$^\circ$ in the laboratory reference frame for an equivalent beam energy of $E_d=21.4$ MeV.  Background from fusion-evaporation on $^{12}$C was accounted for by subtracting data from reactions on a carbon target.}\label{Q}
\end{figure}

The reaction Q-value was calculated on an event-by-event basis.  Fig.~\ref{Q} shows the Q-value spectrum at 140$^{\circ}$ in the lab frame. The energy resolution in Q-value for the (d,p) reaction was more than sufficient to resolve peaks from the population of the ground state and the first excited state at 0.320 MeV.    The background was identified as fusion-evaporation on $^{12}$C by measuring reactions on a carbon target.  It was then possible to account for this background by fitting an exponential curve for the data at E$_d$~=~12.0, 15.0, and 18.0~MeV, and subtracting measured background at E$_d$~=~21.4~MeV.  The background was significantly less obtrusive for curve fitting at the higher energies as the energies of protons from the (d,p) reaction increase more with beam energy than do the energies of fusion-evaporation ejectiles.  

The two bound states in $^{11}$Be were populated at each beam energy.  Data were extracted for the 1.78~MeV resonance from the three higher beam energy measurements.  
The peaks for the bound states were fitted with Gaussian curves.  A Voigt profile was used to fit the peak at 1.78~MeV, using the Gaussian widths (associated with detector resolution) found for the bound states in each spectrum and treating the natural width of the state as a free variable.  
For the data taken with ORRUBA, Gaussian curves were fitted to each peak resulting in a Q-value resolution of  approximately 200~keV.

\begin{figure}
\centering
\includegraphics[width=9.cm]{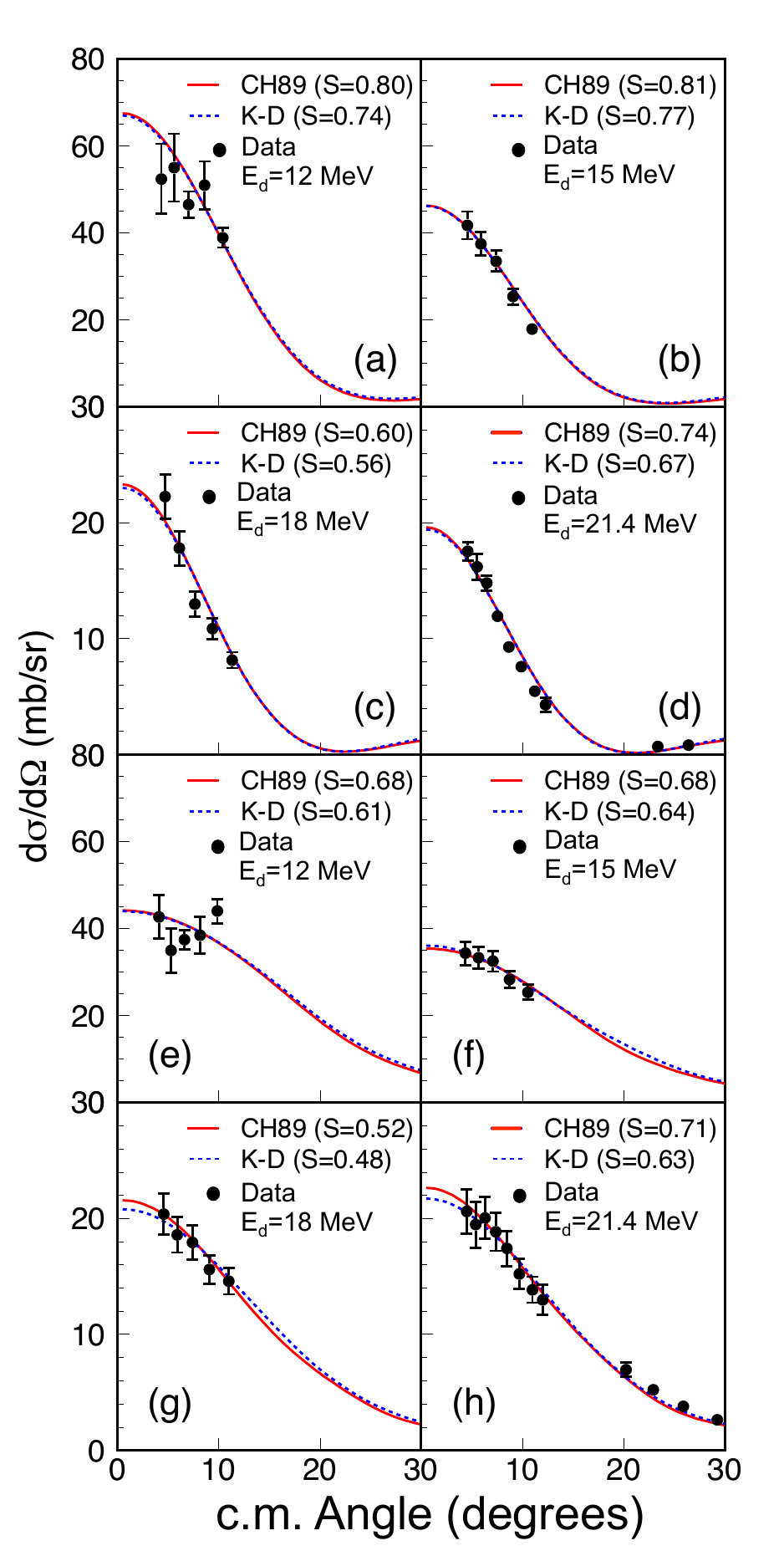}
\caption{(color online).  Differential cross-sections for $^{10}$Be(d,p)$^{11}$Be(gs) for deuteron energies of (a)~12, (b)~15, (c)~18, and (d)~21.4~MeV, and $^{10}$Be$(d,p)^{11}$Be(0.32 MeV) at (e)~12, (f)~15, (g)~18, and (h)~21.4~MeV.  Cross sections were calculated using the adiabatic model of Johnson and Tandy \cite{Joh74}, built using the nucleon potentials of Varner (CH89) \cite{Var91} and Koning and Delaroche \cite{Kon03}.  Calculated cross-sections are scaled using the indicated spectroscopic factors.\label{transfer1}}
\end{figure}

Angular distributions of protons emitted from the $^2$H($^{10}$Be,p)$^{11}$Be reaction to the ground and first excited state are presented in Fig. \ref{transfer1}. The curves show the results of FR-ADWA calculations using the global optical potentials CH89 and K-D as used for the elastic scattering channel above.   The spectroscopic factors, S, were extracted for each state at each energy by scaling the calculation to the data.  The shape of the experimental angular distributions are well reproduced by calculations using either CH89 or K-D, however there is some variance in the intensity of up to 13 \% between the calculations.

Cross-section data for population of the resonance at 1.78~MeV could be extracted at the higher three energies and are presented in Fig.~\ref{resonance}.  The protons emitted following transfer to the 1.78~MeV resonance at E$_d=12$~MeV were too low in energy to extract a reasonable angular distribution.  The transfer to the resonance data will be discussed more fully in the next section.

\section{\label{sec:analysis}Transfer Analysis and Discussion\protect\\ }

The conventional method of analyzing data from transfer reactions using the Distorted Wave Born Approximation (DWBA) has been shown to be particularly sensitive to the optical potentials used \cite{Sch12}.   In the 1970s, Johnson and Soper \cite{Joh72} showed the importance of including the breakup channel for the deuteron in the theoretical treatment of (d,p) reactions, and by using a zero-range formulated a method called the ADiabatic Wave Approximation (ADWA) . The finite-range version by Johnson and Tandy \cite{Joh74} (FR-ADWA) was recently applied to a wide range of reactions by Nguyen, Nunes, and Johnson \cite{Ngu10}.  

The transfer data presented are analyzed using FR-ADWA. In FR-ADWA, the reaction is treated as a three-body n+p+$^{10}$Be problem, and thus neutron and proton optical potentials are needed, along with the binding potentials for the deuteron and the final state in $^{11}$Be. For the nucleon global potentials we take CH89 and K-D, shown in Fig.\ref{pelasticplot} to reproduce the elastic scattering fairly well. The single-particle parameters used for the overlap function for $^{11}$Be were: radius $r~=~1.25$~fm, diffuseness  $a~=~0.65$~fm and a spin-orbit term with $V_{so}=5.5$~MeV and the same geometry as the central interaction. The central depth was adjusted to reproduce the neutron separation energy of the desired state. 
 As in previous work, the Reid  interaction \cite{Rei68} was taken for the deuteron bound state and the transfer operator. The effective adiabatic potential was computed with {\sc twofnr} \cite{twofnr} and the transfer calculations were performed with {\sc fresco} \cite{Tho88}.

The spectroscopic factors (S) extracted from the angular distributions using the CH89 and K-D optical potentials are presented in Table \ref{sfsgs} for population of the ground state and Table \ref{sfs32} for the bound excited state.  They are shown graphically in Fig. \ref{sf} with error boxes centered on the mean value for each state with each potential over the four energies.  With the exception of the measurement at E$_d$~=~18~MeV the average spectroscopic factor agrees with the individual measurements within the limit of the uncertainties.  The average S extracted with CH89 also agrees with that from K-D.  The 18 MeV (d,p) data appear to be systematically lower than at other energies.  The normalization of these data, the analysis and the experimental setup including the target were the same as those used in the 12-MeV measurement and therefore we feel these data are reliable.  With four data points it should be expected that one should fall outside the one sigma error bar.

Figure~\ref{sf_all} shows the average S extracted from the current data using CH89 and K-D optical potentials compared to the literature values tabulated in table~\ref{spec}.  The current value of S for the ground state of $^{11}$Be is either in agreement with, or close to agreement with the earlier transfer measurements \cite{Aut70,Zwi79} as well as most of the breakup and Coulomb dissociation measurements \cite{Aum00, Pal03, Fuk04}.  The value of S for the ground state extracted by Lima et al. \cite{Lim07}, despite its relatively large error bar, is lower than the value presented here.  The calculation by Vinh Mau \cite{Vin95} using vibrational couplings significantly overestimates the value of S for the ground state, whereas that of Nunes, Thompson, and Johnson \cite{Nun96} using rotational couplings gives a value in agreement with the current value.  

There are fewer measurements related to the first excited state, and here the current value agrees with that from Auton et al \cite{Aut70}.  This agreement was not necessarily expected since the normalization procedure in the work of reference \cite{Aut70} was subject to significant systematic uncertainties. The current value of S for the first excited state is lower than that extracted from the transfer measurement of Zwieglinski \cite{Zwi79} and the two theoretical values used here for comparison \cite{Vin95,Nun96}. 

\begin{figure}
\centering
\includegraphics[width=10.cm]{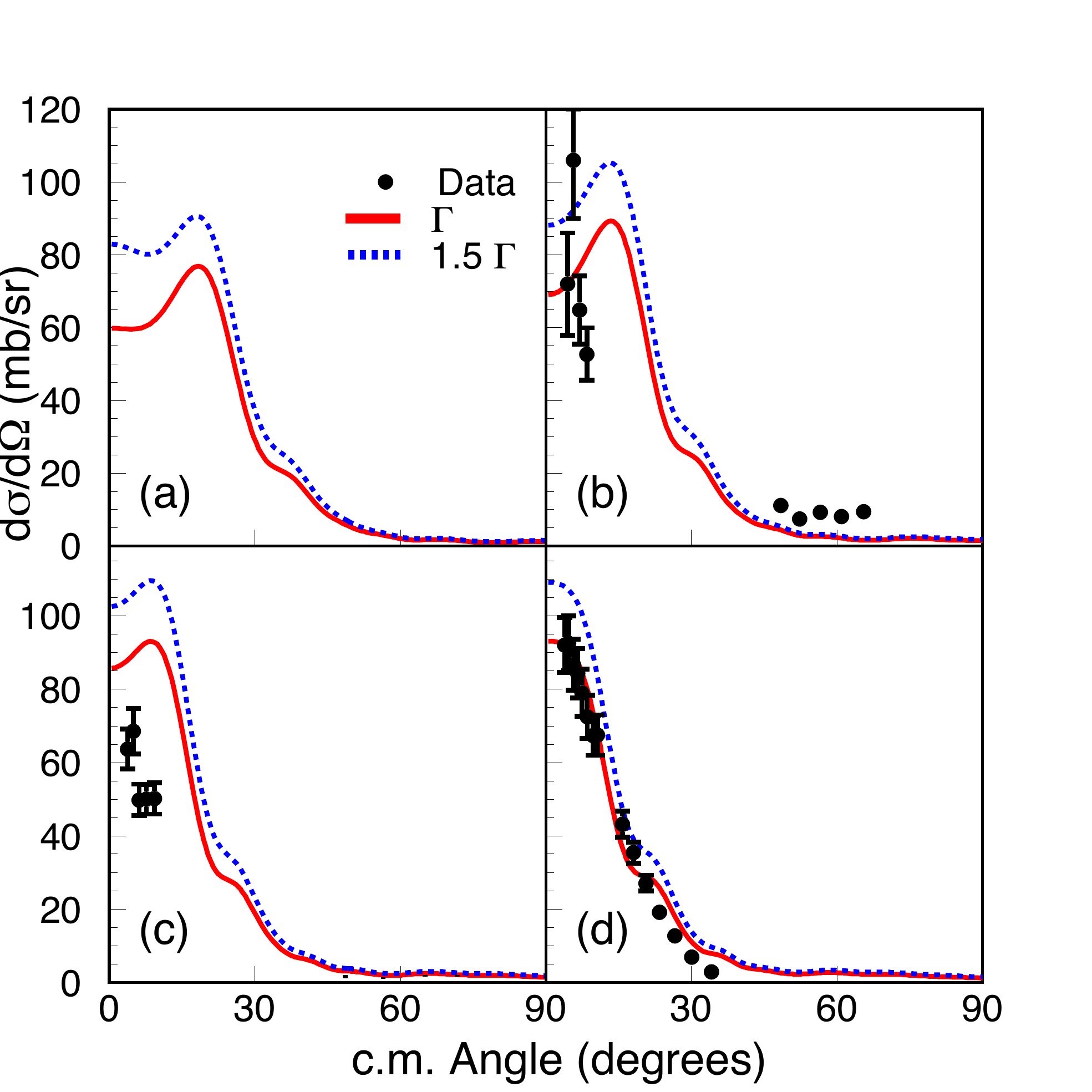}
\caption{(color online).  Differential cross-sections are presented for transfer to the first resonance in $^{11}$Be at 1.78~MeV via the $^{10}$Be(d,p) reaction in inverse kinematics at deuteron energies of (a)~12, (b)~15, (c)~18, and (d)~21.4~MeV.  The curves are from FR-ADWA calculations using (solid line) an energy bin that is the same width as for the resonance used in the calculation and (dotted line) with a width 1.5 times that value.  At 12~MeV the protons were too low in energy to extract an angular distribution.}\label{resonance}.  
\end{figure}

\begin{figure}[tr]
\centering
\includegraphics[trim=1mm 10mm 10mm 10mm, clip, width=8.0cm]{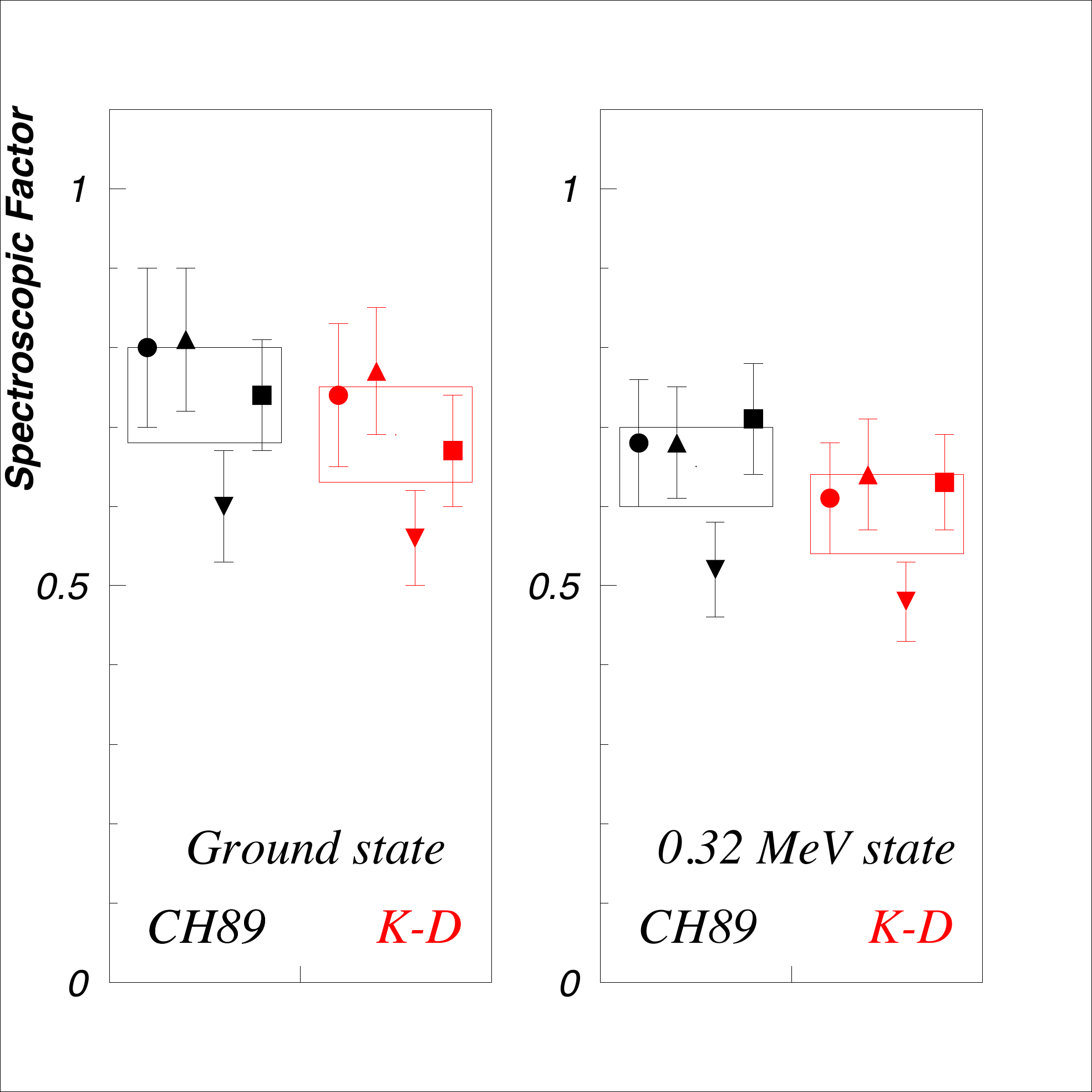}
\caption{(color online).  Spectroscopic factors extracted from the current data using the optical potentials of Varner (CH89)\cite{Var91} and Koning and Delaroche (K-D)\cite{Kon03}.  The points are for spectroscopic factors extracted from the data at equivalent deuteron energies of  12.0 (circles), 15.0 (triangles), 18.0 (inverted triangles), and 21.4 (squares) MeV.  The boxes are centered on the average and show the uncertainty on the average value.  \label{sf}}
\end{figure}

\begin{figure}
\centering
\includegraphics[trim=1mm 10mm 10mm 10mm, clip, width=8.0cm]{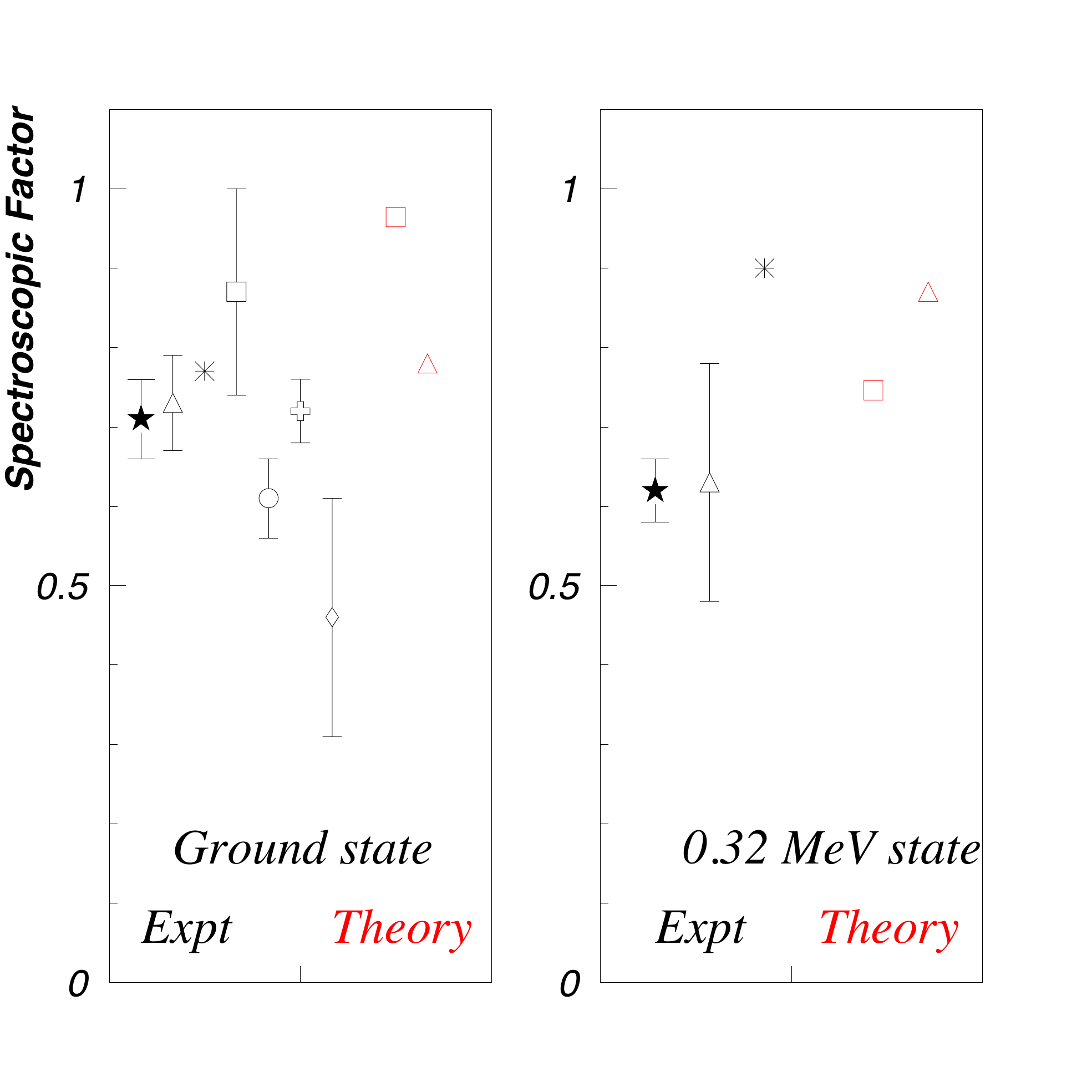}
\caption{(color online).  Spectroscopic factors extracted from the current data (filled stars) compared to previous measurements (black) and theory (red) for the ground and first excited states. The comparison points are the same works as those included in table \ref{spec}, from references \cite{Aut70} (triangles), \cite{Zwi79} (asterixes), \cite{Aum00} (square), \cite{Pal03}(circle), \cite{Fuk04} (open cross), \cite{Lim07} (open diamond), \cite{Vin95} (red squares), and \cite{Nun96} (red triangles).   \label{sf_all}}
\end{figure}

Transfer to the d$_{5/2}$ resonance at 1.78~MeV was calculated \cite{Bey12} in a similar manner to transfer to the bound states described above, using the FR-ADWA framework and the CH89 optical potential.  The n-$^{10}$Be potential depths, both central and spin orbit, were adjusted to reproduce the resonance energy.  This resulted in a real width of the resonance of 0.192~MeV, almost twice the measured value of 0.1(0.01)~MeV \cite{Hai09, Kel12}.  The final state was described in the calculation as an energy bin with the width set to the same size as that used in the model for the resonance, such that all the strength of the resonance could be captured.  In principle, the calculation should be insensitive to the size of this fictitious energy bin.  To test this, the calculation was repeated with the bin set to be 50~\% larger than the width used for the resonance.  The results of these calculations are compared to the data in Fig.~\ref{resonance}.  The sensitivity of the calculation to the width of the energy bin precludes the extraction of spectroscopic factors in this case.

\begin {table}
\caption{\label{sfsgs}
Ground state spectroscopic factors extracted for each set of optical model parameters in the exit channel.  Uncertainties include only experimental contributions.
}
\begin{ruledtabular}
\begin{tabular}{ccc} 
\textrm{E$_d$ (MeV)}&\textrm{CH89}&\textrm{K-D}\\
\colrule
   12 	&	0.80$\pm$0.10	&	0.74$\pm$0.09\\
   15 	&	0.81$\pm$0.09	&	0.77$\pm$0.08\\
   18 	&	0.60$\pm$0.07	&	0.56$\pm$0.06\\
   21.4 	&	0.74$\pm$0.07	&	0.67$\pm$0.07\\
\colrule
   \textrm{Average}	&	0.74$\pm$0.06	&	0.69$\pm$0.06	\\
\end{tabular}
\end{ruledtabular}
\end{table}

\begin {table}
\caption{\label{sfs32}
First excited state spectroscopic factors extracted for each set of optical model parameters in the exit channel.  Uncertainties include only experimental contributions.
}
\begin{ruledtabular}
\begin{tabular}{ccc} 
\textrm{E$_d$ (MeV)}&\textrm{CH89}&\textrm{K-D}\\
\colrule
   12 	&	0.68$\pm$0.08	&	0.61$\pm$0.07\\
   15 	&	0.68$\pm$0.07	&	0.64$\pm$0.07\\
   18 	&	0.52$\pm$0.06	&	0.48$\pm$0.05\\
   21.4 	&	0.71$\pm$0.07	&	0.63$\pm$0.06\\
\colrule
   \textrm{Average}	&	0.65$\pm$0.05	&	0.59$\pm$0.05	\\
\end{tabular}
\end{ruledtabular}
\end{table}

\section{\label{sec:conclusions}Conclusions:\protect\\ }
Data are presented here for elastic scattering of $^{10}$Be on protons and deuterons, inelastic scattering on deuterons, and one-neutron transfer to the two bound states of $^{11}$Be at four energies ranging from E$_d$~=~12 to 21.4~MeV, all in inverse kinematics.  Additionally, transfer to the first resonance at 1.78~MeV was measured for the higher three energies.  Absolute cross sections, over the range of angles covered, were measured in all cases.  

For the proton scattering data, no single potential could reproduce the data at all four energies.  However, the CH89 and K-D potentials performed generally better than the older Wat potential.  The transfer calculations presented here therefore used the CH89 and K-D potentials.  In order to simultaneously reproduce the elastic and inelastic scattering on deuterons, a new potential was fitted to the data, starting from P-P and a nuclear deformation length of $\delta_2=1.84$~fm.  The depth of the real part of these potentials is significantly smaller and the radius larger than typical for stable nuclei, as is common for effective potentials extracted in this way for weakly-bound systems.  Calculations using these new potentials provided reasonable agreement with the elastic data, and the inelastic data, except at the highest energy.  At E$_d=21.4$~MeV other channels need to be taken into account in the calculation.

The transfer to bound states in $^{11}$Be show a level of consistency in the extracted spectroscopic factors that can provide confidence in the use of FR-ADWA for these data.  The values of 0.71(5) and 0.62(4) for the spectroscopic factor of the ground and first excited states can be used to guide future theoretical nuclear structure studies.

The calculations for the first resonant state showed sensitivity to the size of the energy bin placed in the continuum, and therefore were not used to extract a spectroscopic factor in this case.  An appropriate theory for extracting detailed structure in the case of unbound states is desirable for future studies.

The data presented here represent the most complete dataset on reactions with a $^{10}$Be beam that are currently available.  They have been analyzed within the optical model and FR-ADWA frameworks and have shown the strengths and limitations of each method.  It is anticipated that future theoretical work on the neutron-rich beryllium isotopes will benefit from the data reported here.

\begin{acknowledgments}
This work was supported by the US Department of Energy
under contract numbers DEFG02-96ER40955 and  DE-FG02-96ER40990 (TTU), DE-AC05-00OR22725 (ORNL), DE-FG03-93ER40789 (Colorado School of Mines),
DE-FG02-96ER40983 and DE-SC0001174 (UT),
DE-FG52-08NA28552 and DE-AC02-06CH11357
(MSU), the National Science Foundation under contract numbers
NSF-PHY0354870 and NSF-PHY0757678 (Rutgers), NSF-PHY-0969456(Notre Dame), and NSF-PHY-0555893
(MSU), the NRF of Korea (MSIP) under contract numbers NRF-2012M7A1A2055625 (Ewha) and NRF-2012R1A1A1041763 (Sungkyunkwan), and the UK Science and Technology Funding Council under contract
number PP/F000715/1.  This research was sponsored in part by the National Nuclear Security Administration 
under the Stewardship Science Academic Alliances program through DOE Cooperative Agreements DE-FG52-03NA00143 and 
DE-FG52-08NA28552 (Rutgers, ORAU).  The authors would like to thank the operations staff at the Holifield Radioactive Ion Beam Facility for making these measurements possible.
\end{acknowledgments}
\bibliography{11Be_PRC}
\include{supplemental-1}

\end{document}

%% file: supplemental-1.tex
\section{Data}
Tabulated data are included for the $^{10}$Be$(p,p)$, $^{10}$Be$(d,d)$, $^{10}$Be$(d,p)$, and  $^{10}$Be$(d,d')$ reactions in inverse kinematics.  Differential cross-sections for elastic scattering are quoted as a ratio to Rutherford scattering cross-sections.  The uncertainties shown are statistical.\\
\begin{table}[h]
\centering
\caption{\label{pp60}
Data from the $^{10}$Be(p,p) reaction in inverse kinematics at E$_p$~=~6~MeV.}
\begin{tabular}{cc}
\hline
\hline
\textrm{$\theta _p (deg.)$}&
\textrm {Ratio to Rutherford cross section}\\
\hline
28.9 & 2.1$\pm$ 0.2 \\
30.1 & 2.3$\pm$ 0.2 \\
31.3 & 2.4$\pm$ 0.1 \\
32.5 & 2.7$\pm$ 0.1 \\
33.6 & 2.8$\pm$ 0.1 \\
34.8 & 2.8$\pm$ 0.1 \\
35.9 & 2.9$\pm$ 0.1 \\
37.0 & 3.3$\pm$ 0.1 \\
38.2 & 3.3$\pm$ 0.1 \\
39.3 & 3.8$\pm$ 0.1 \\
40.6 & 3.8$\pm$ 0.1 \\
42.0 & 4.2$\pm$ 0.2 \\
43.1 & 3.8$\pm$ 0.2 \\
44.2 & 4.3$\pm$ 0.2 \\
45.3 & 4.3$\pm$ 0.2 \\
46.4 & 4.2$\pm$ 0.2 \\
47.4 & 4.6$\pm$ 0.2 \\
48.5 & 4.5$\pm$ 0.2 \\
49.5 & 5.2$\pm$ 0.2 \\
50.5 & 4.6$\pm$ 0.2 \\
51.8 & 4.5$\pm$ 0.2 \\
53.4 & 4.7$\pm$ 0.2 \\
54.7 & 4.6$\pm$ 0.2 \\
56.0 & 4.7$\pm$ 0.3 \\
\hline
\hline
\end{tabular}
\end{table}

\begin{table}[h]
\centering
\caption{\label{tpp75}
Data from the $^{10}$Be(p,p) reaction in inverse kinematics at E$_p$~=~7.5 MeV.}
\begin{tabular}{cc}
\hline
\hline
\textrm{$\theta _p (deg.)$}&
\textrm {Ratio to Rutherford cross section}\\
\hline
27.8 & 	2.3 $\pm$ 0.05 \\
28.9 & 	2.8 $\pm$ 0.04 \\
30.1 & 	3.3 $\pm$ 0.05 \\
31.3 & 	3.6 $\pm$ 0.06 \\
32.5 & 	3.9 $\pm$ 0.06 \\
33.6 & 	4.2 $\pm$ 0.07 \\
34.8 & 	4.5 $\pm$ 0.07 \\
35.9 & 	4.7 $\pm$ 0.08 \\
37.0 & 	5.0 $\pm$ 0.09 \\
38.2 & 	5.2 $\pm$ 0.10 \\
39.3 & 	5.5 $\pm$ 0.10 \\
40.6 & 	5.8 $\pm$ 0.09 \\
42.0 & 	6.1 $\pm$ 0.12 \\
43.1 & 	6.4 $\pm$ 0.13 \\
44.2 & 	6.5 $\pm$ 0.14 \\
45.3 & 	6.0 $\pm$ 0.14 \\
46.4 & 	6.6 $\pm$ 0.2 \\
47.4 & 	6.3 $\pm$ 0.2 \\
48.5 & 	6.8 $\pm$ 0.2 \\
49.5 & 	6.7 $\pm$ 0.2 \\
50.5 & 	6.4 $\pm$ 0.2 \\
51.8 & 	6.1 $\pm$ 0.2 \\
53.4 & 	6.1 $\pm$ 0.2 \\
54.7 & 	5.9 $\pm$ 0.2 \\
56.0 & 	5.9 $\pm$ 0.2 \\
\hline
\hline
\end{tabular}
\end{table}
\begin{table}[h]
\centering
\caption{\label{pp90}
Data from the $^{10}$Be(p,p) reaction in inverse kinematics at E$_p$~=~9 MeV.}
\begin{tabular}{cc}
\hline
\hline
\textrm{$\theta _p (deg.)$}&
\textrm {Ratio to Rutherford cross section}\\
\hline
27.8 & 	3.5$\pm$0.09 \\
28.9 & 	4.0$\pm$0.08 \\
30.1 & 	4.4$\pm$0.09 \\
31.3 & 	4.9$\pm$0.10 \\
32.5 & 	5.1$\pm$0.11 \\
33.6 & 	5.7$\pm$0.12 \\
34.8 & 	5.8$\pm$0.13 \\
35.9 & 	6.1$\pm$0.14 \\
37.0 & 	6.4$\pm$0.2 \\
38.2 & 	7.0$\pm$0.2 \\
39.3 & 	7.1$\pm$0.2 \\
40.6 & 	7.6$\pm$0.2 \\
42.0 & 	7.7$\pm$0.2 \\
43.1 & 	7.8$\pm$0.2 \\
44.2 & 	7.6$\pm$0.2 \\
45.3 & 	8.2$\pm$0.3 \\
46.4 & 	8.3$\pm$0.3 \\
47.4 & 	8.1$\pm$0.3 \\
48.5 & 	8.5$\pm$0.3 \\
49.5 & 	8.9$\pm$0.3 \\
50.5 & 	8.3$\pm$0.3 \\
51.8 & 	8.2$\pm$0.3 \\
53.4 & 	7.7$\pm$0.3 \\
54.7 & 	7.5$\pm$0.3 \\
56.0 & 	7.3$\pm$0.4 \\
\hline
\hline
\end{tabular}
\end{table}
\begin{table}[h]
\centering
\caption{\label{pp107}
Data from the $^{10}$Be(p,p) reaction in inverse kinematics at E$_p$~=~10.7 MeV.}
\begin{tabular}{cc}
\hline
\hline
\textrm{$\theta _p (deg.)$}&
\textrm {Ratio to Rutherford cross section}\\
\hline
26.5 & 	4.3$\pm$   0.11 \\
27.8 & 	5.1$\pm$   0.12 \\
28.9 & 	5.4$\pm$   0.10 \\
30.1 & 	6.0$\pm$   0.11 \\
31.3 & 	6.6$\pm$   0.13 \\
32.5 & 	6.8$\pm$   0.14 \\
33.6 & 	7.1$\pm$   0.15 \\
34.8 & 	7.8$\pm$   0.2 \\
35.9 & 	8.0$\pm$   0.2 \\
37.0 & 	8.8$\pm$   0.2 \\
38.2 & 	9.0$\pm$   0.2 \\
39.3 & 	8.9$\pm$   0.2 \\
40.6 & 	9.5$\pm$   0.2 \\
42.0 & 	9.5$\pm$   0.3 \\
43.1 & 	10.0$\pm$  0.3 \\
44.2 & 	10.1$\pm$  0.3 \\
45.3 & 	9.4$\pm$   0.3 \\
46.4 & 	9.5$\pm$   0.3 \\
47.4 & 	9.3$\pm$   0.3 \\
48.5 & 	9.4$\pm$   0.4 \\
49.5 & 	9.2$\pm$   0.4 \\
50.5 & 	9.4$\pm$   0.4 \\
51.8 & 	7.7$\pm$   0.3 \\
53.4 & 	8.2$\pm$   0.4 \\
54.7 & 	7.1$\pm$   0.5 \\
56.0 & 	6.1$\pm$   0.6 \\
\hline
\hline
\end{tabular}
\end{table}
\begin{table}[h]
\centering
\caption{\label{tab1}
Data from the $^{10}$Be(d,d) reaction in inverse kinematics at E$_d$~=~12 MeV.}
\begin{tabular}{cc}
\hline
\hline
\textrm{$\theta _d (deg.)$}&
\textrm {Ratio to Rutherford cross section}\\
\hline
   33.2     &    5.74 $\pm$ 0.15    \\
   35.7     &    4.44 $\pm$ 0.19    \\
   38.1     &    2.69 $\pm$ 0.18   \\ 
   40.4     &    1.40 $\pm$ 0.15   \\ 
   43.8     &   0.12 $\pm$ 0.10  \\
   47.0     &    1.23 $\pm$ 0.13    \\
   50.2     &    2.67 $\pm$ 0.30    \\
   53.4     &    4.17 $\pm$ 0.25    \\
   56.6     &    4.76 $\pm$ 0.30    \\
   59.8     &    5.09 $\pm$ 0.21    \\
   63.0     &    5.02 $\pm$ 0.17    \\
   66.2     &    5.45 $\pm$ 0.31    \\
\hline
\hline
\end{tabular}
\centering
\caption{\label{tab2}
Data from the  $^{10}$Be(d,d) reaction in inverse kinematics at E$_d$~=~15 MeV. }
\begin{tabular}{cc}
\hline
\hline
\textrm{$\theta _d (deg.)$}&
\textrm {Ratio to Rutherford cross section} \\
\hline
   29.6     &    8.40 $\pm$ 0.13    \\
   31.8     &    5.88 $\pm$ 0.11    \\
   33.9     &    4.13 $\pm$ 0.11    \\
   36.0     &    2.67 $\pm$ 0.12    \\
   37.9     &    1.45 $\pm$ 0.10   \\
   39.7     &   0.46 $\pm$ 0.10    \\
   40.6     &   0.23 $\pm$ 0.07    \\
   43.8     &   0.71 $\pm$ 0.18    \\
   47.0     &    2.64 $\pm$ 0.16    \\
   50.2     &    4.25 $\pm$ 0.21    \\
   53.4     &    5.42 $\pm$ 0.25    \\
   56.5     &    6.55 $\pm$ 0.51    \\
   59.7     &    6.40 $\pm$ 0.60    \\
\hline
\hline
\end{tabular}
\centering
\caption{\label{tab3}
Data from the $^{10}$Be(d,d) reaction in inverse kinematics at E$_d$~=~18 MeV.
}
\begin{tabular}{cc}
\hline 
\hline
\textrm{$\theta _d (deg.)$}&
\textrm {Ratio to Rutherford cross section} \\
\hline
   29.0    &     5.67 $\pm$ 0.19    \\
   30.9    &     3.58 $\pm$ 0.22    \\
   32.7    &     2.33 $\pm$ 0.26    \\
   34.5    &     1.73 $\pm$ 0.30    \\
   37.4    &    0.43 $\pm$ 0.20    \\
   40.6    &    0.71 $\pm$ 0.16    \\
   43.7    &     1.96 $\pm$ 0.22    \\
   46.9    &     3.71 $\pm$ 0.23    \\
   50.1    &     4.25 $\pm$ 0.27    \\
   53.3    &     4.92 $\pm$ 0.33    \\
   56.5    &     3.56 $\pm$ 0.23    \\
   59.7    &     3.87 $\pm$ 0.59      \\
\hline 
\hline
\end{tabular}
\end{table}
\begin{table}
\centering
\caption{\label{tab4}
Data from the $^{10}$Be(d,d) reaction in inverse kinematics at E$_d$~=~21.4 MeV. 
}
\begin{tabular}{cc}
\hline 
\hline
\textrm{$\theta _d (deg.)$}&
\textrm {Ratio to Rutherford cross section} \\
\hline
30.8	&	0.18 $\pm$ 0.02\\
32.8	&	0.36 $\pm$ 0.03\\
34.7	&	0.82 $\pm$ 0.05\\
36.7	&	1.85 $\pm$ 0.07\\
38.7	&	2.94 $\pm$ 0.09\\
40.7	&	4.17 $\pm$ 0.11\\
42.4	&	4.86 $\pm$ 0.13\\
44.7	&	5.90 $\pm$ 0.15\\
46.7	&	5.91 $\pm$ 0.17\\
48.7	&	5.88 $\pm$ 0.18\\
50.7	&	5.47 $\pm$ 0.18\\
52.6	&	4.42 $\pm$ 0.18\\
54.6	&	3.51 $\pm$ 0.18\\
56.6	&	2.81 $\pm$ 0.17\\
58.6	&	1.20 $\pm$ 0.21\\
\hline 
\hline
\end{tabular}
\centering
\caption{\label{tab5}
Data from the $^{10}$Be(d,p) reaction in inverse kinematics populating the ground state of $^{11}$Be at E$_d$~=~12~MeV.}
\begin{tabular}{cc}
\hline
\hline
\textrm{$\theta _p (deg.)$}&
\textrm {Differential cross section (mb/sr)} \\
\hline
   4.3     &    52.4 $\pm$ 8.1    \\
   5.6     &    55.0 $\pm$ 7.8   \\
   7.0     &    46.4 $\pm$ 3.0    \\
   8.6     &    50.9 $\pm$ 5.5    \\
   10.4    &     38.8 $\pm$ 2.2    \\
   53.0    &     4.09 $\pm$ 0.52    \\
   56.8    &     2.31 $\pm$ 0.26    \\
   60.9    &     1.92 $\pm$ 0.27    \\
   65.1    &     1.57 $\pm$ 0.21    \\
   69.5    &    0.95 $\pm$ 0.18    \\
   74.0    &    0.64 $\pm$ 0.24    \\
   78.6    &    0.82 $\pm$ 0.23    \\
\hline
\hline
\end{tabular}
\centering
\caption{\label{tab6}
Data from the $^{10}$Be(d,p) reaction in inverse kinematics populating the ground state of $^{11}$Be at E$_d$~=~15~MeV.}
\begin{tabular}{cc}
\hline
\hline
\textrm{$\theta _p (deg.)$}&
\textrm {Differential cross section (mb/sr)} \\
\hline
   4.6     &    41.7 $\pm$ 3.2\\
   5.9     &    37.4 $\pm$ 2.7  \\  
   7.4     &    33.4 $\pm$ 2.4    \\
   9.1     &    25.3 $\pm$ 1.8    \\
   11.0     &    17.9 $\pm$ 1.3   \\ 
   58.0     &    4.17 $\pm$ 0.69 \\   
   62.0     &    2.99 $\pm$ 0.40   \\ 
   66.1     &    1.49 $\pm$ 0.37    \\
   70.5     &    0.27 $\pm$ 0.23    \\
\hline
\hline
\end{tabular}
\end{table}
\begin{table}
\centering
\caption{\label{tab7}
Data from the $^{10}$Be(d,p) reaction in inverse kinematics populating the ground state of $^{11}$Be at E$_d$~=~18~MeV.
}
\begin{tabular}{cc}
\hline
\hline
\textrm{$\theta _p (deg.)$}&
\textrm {Differential cross section (mb/sr)} \\
\hline
   4.7     &    22.2 $\pm$ 1.9\\
   6.1     &    17.8 $\pm$ 1.5\\
   7.6     &    13.0 $\pm$ 1.1  \\  
   9.4     &    10.8 $\pm$ 0.9\\    
   11.3    &     8.13 $\pm$ 0.69  \\  
   53.0    &    0.78 $\pm$ 0.22    \\
   60.7    &    0.74 $\pm$ 0.17    \\
   69.0    &    0.23 $\pm$ 0.11    \\
\hline
\hline
\end{tabular}
\centering
\caption{\label{tab8}
Data from the $^{10}$Be(d,p) reaction in inverse kinematics populating the ground state of $^{11}$Be at E$_d$~=~21.4~MeV.
}
\begin{tabular}{cc}
\hline
\hline
\textrm{$\theta _p (deg.)$}&
\textrm {Differential cross section (mb/sr)} \\
\hline
4.6 &	17.52 $\pm$ 0.81\\
5.5 &	16.18 $\pm$ 1.09\\
6.5 &	14.78 $\pm$ 0.65\\
7.5 &	11.93 $\pm$ 0.47\\
8.7 &	9.26 $\pm$ 0.42\\
9.9 &	7.55 $\pm$ 0.34\\
11.2 &	5.44 $\pm$ 0.28\\
12.3 &	4.28 $\pm$ 0.61\\
18.2 &	0.85 $\pm$ 0.18\\
20.7 &	0.13 $\pm$ 0.13\\
23.4 &	0.68 $\pm$ 0.13\\
26.4 &	0.81 $\pm$ 0.11\\
29.8 &	1.21 $\pm$ 0.09\\
33.5 &	1.19 $\pm$ 0.08\\
37.6 &	0.65 $\pm$ 0.05\\
\hline
\hline
\end{tabular}
\centering
\caption{\label{tab9}
Data from the $^{10}$Be(d,p) reaction in inverse kinematics populating the first excited state of $^{11}$Be at E$_{ex}$~=~0.320 MeV at E$_d$~=~12~MeV.}
\begin{tabular}{cc}
\hline
\hline
\textrm{$\theta _p (deg.)$}&
\textrm {Differential cross section (mb/sr)} \\
\hline
   4.1    &     42.6 $\pm$ 5.0    \\
   5.3    &     34.9 $\pm$ 5.2    \\
   6.6    &     37.4 $\pm$ 2.2    \\
   8.1    &     38.4 $\pm$ 4.2    \\
   9.9    &     44.0 $\pm$ 2.8    \\
   51.8    &     5.41 $\pm$ 0.61    \\
   55.7    &     4.45 $\pm$ 0.33    \\
   59.7    &     3.91 $\pm$ 0.34    \\
   64.0    &     3.43 $\pm$ 0.27    \\
   68.4    &     2.65 $\pm$ 0.24    \\
   73.0    &     2.02 $\pm$ 0.32    \\
   77.7    &     1.50 $\pm$ 0.34    \\
\hline
\hline
\end{tabular}
\centering
\caption{\label{tab10}
Data from the $^{10}$Be(d,p) reaction in inverse kinematics populating the first excited state of $^{11}$Be at E$_{ex}$~=~0.320 MeV at E$_d$~=~15~MeV.}
\begin{tabular}{cc}
\hline
\hline
\textrm{$\theta _p (deg.)$}&
\textrm {Differential cross section (mb/sr)} \\
\hline
   4.4     &    34.3 $\pm$ 2.7    \\
   5.7     &    33.2 $\pm$ 2.5    \\
   7.1     &    32.4 $\pm$ 2.4    \\
   8.7     &    28.3 $\pm$ 2.0    \\
   10.6   &     25.3 $\pm$ 1.8    \\
   57.1   &     4.85 $\pm$ 0.75    \\
   61.1   &     4.48 $\pm$ 0.46    \\
   65.3   &     3.22 $\pm$ 0.48    \\
   69.7   &     2.18 $\pm$ 0.35    \\
\hline
\hline
\end{tabular}
\end{table}
\begin{table}
\centering
\caption{\label{tab11}
Data from the $^{10}$Be(d,p) reaction in inverse kinematics populating the first excited state of $^{11}$Be at E$_{ex}$~=~0.320 MeV at E$_d$~=~18~MeV.}
\begin{tabular}{cc}
\hline
\hline
\textrm{$\theta _p (deg.)$}&
\textrm {Differential cross section (mb/sr)} \\
\hline
   4.6     &    20.4 $\pm$ 1.8    \\
   5.9     &    18.6 $\pm$ 1.5    \\
   7.4     &    17.9 $\pm$ 1.5    \\
   9.1     &    15.6 $\pm$ 1.2    \\
   11.0    &    14.6 $\pm$ 1.2    \\
   52.3    &    1.34 $\pm$ 0.28    \\
   60.0    &    0.85 $\pm$ 0.18    \\
   68.3    &    0.37 $\pm$ 0.13    \\
\hline
\hline
\end{tabular}
\centering
\caption{\label{tab12}
Data from the $^{10}$Be(d,p) reaction in inverse kinematics populating the first excited state of $^{11}$Be at E$_{ex}$~=~0.320 MeV at E$_d$~=~21.4~MeV.}
\begin{tabular}{cc}
\hline
\hline
\textrm{$\theta _p (deg.)$}&
\textrm {Differential cross section (mb/sr)} \\
\hline
4.5 &	20.6 $\pm$ 1.9\\
5.4 &	19.5 $\pm$ 2.0\\
6.3 &	20.0 $\pm$ 1.8\\
7.4 &	18.8 $\pm$ 1.6\\
8.5 &	17.4 $\pm$ 1.5\\
9.7 &	15.2 $\pm$ 1.3\\
11.0 &	13.8 $\pm$ 1.1\\
12.0 &	13.0 $\pm$ 1.3\\
17.8 &	9.14 $\pm$ 0.78\\
20.2 &	6.93 $\pm$ 0.60\\
22.9 &	5.23 $\pm$ 0.45\\
25.9 &	3.78 $\pm$ 0.33\\
29.2 &	2.61 $\pm$ 0.23\\
32.9 &	2.23 $\pm$ 0.20\\
37.0 &	2.16 $\pm$ 0.19\\
\hline
\hline
\end{tabular}
\centering
\caption{\label{tab13}
Data from the $^{10}$Be(d,p) reaction in inverse kinematics populating the resonance of $^{11}$Be at E$_{ex}$~=~1.78 MeV at E$_d$~=~15~MeV.}
\begin{tabular}{cc}
\hline
\hline
\textrm{$\theta _p (deg.)$}&
\textrm {Differential cross section (mb/sr)} \\
\hline
   4.5     &    72. $\pm$ 14    \\
   5.6     &    106. $\pm$ 16    \\
   6.9     &    64.8 $\pm$ 9.4    \\
   8.4     &    52.7 $\pm$ 7.2   \\
   48.4    &     11.1 $\pm$ 0.9   \\ 
   52.3    &     7.36 $\pm$ 0.74    \\
   56.5    &     9.24 $\pm$ 0.76    \\
   60.9    &     8.05 $\pm$ 0.76    \\
   65.5    &     9.43 $\pm$ 0.74  \\
\hline
\hline
\end{tabular}
\centering
\caption{\label{tab14}
Data from the $^{10}$Be(d,p) reaction in inverse kinematics populating the first excited state of $^{11}$Be at E$_{ex}$~=~1.78 MeV at E$_d$~=~18~MeV.}
\begin{tabular}{cc}
\hline
\hline
\textrm{$\theta _p (deg.)$}&
\textrm {Differential cross section (mb/sr)} \\
\hline
   3.8     &    63.7 $\pm$ 5.4    \\
   5.0     &    68.5 $\pm$ 6.2    \\
   6.2     &    49.8 $\pm$ 4.3    \\
   7.7     &    50.0 $\pm$ 4.0    \\
   9.3     &    50.2 $\pm$ 4.3    \\
   48.6    &     1.63 $\pm$ 0.47    \\
   56.5    &     1.28 $\pm$ 0.22    \\
   65.1    &     1.87 $\pm$ 0.28    \\
   71.8    &     1.85 $\pm$ 0.30    \\
\hline
\hline
\end{tabular}
\end{table}
\begin{table}
\centering
\caption{
Data from the $^{10}$Be(d,p) reaction in inverse kinematics populating the first excited state of $^{11}$Be at E$_{ex}$~=~1.78 MeV at E$_d$~=~21.4~MeV.}\label{tab15}
\begin{tabular}{cc}
\hline
\hline
\textrm{$\theta _p (deg.)$}&
\textrm {Differential cross section (mb/sr)} \\
\hline
3.9 &	92.0 $\pm$ 7.5\\
4.7 &	92.5 $\pm$ 7.5\\
5.6 &	86.6 $\pm$ 7.0\\
6.5 &	84.3 $\pm$ 6.8\\
7.4 &	79.0 $\pm$ 6.4\\
8.5 &	72.5 $\pm$ 5.9\\
9.7 &	67.3 $\pm$ 5.4\\
10.6 &	67.5 $\pm$ 5.5\\
15.8 &	43.2 $\pm$ 3.5\\
18.1 &	35.5 $\pm$ 2.9\\
20.6 &	27.1 $\pm$ 2.2\\
23.4 &	19.2 $\pm$ 1.6\\
26.6 &	12.7 $\pm$ 1.0\\
30.1 &	7.05 $\pm$ 0.57\\
34.1 &	2.92 $\pm$ 0.24\\
\hline
\hline
\end{tabular}
\centering
\caption{\label{tab16}
Data from the $^{10}$Be(d,d') reaction in inverse kinematics populating the first excited state of $^{10}$Be at E$_{ex}$~=~3.368 MeV at E$_d$~=~12 MeV.}
\begin{tabular}{cc}
\hline
\hline
\textrm{$\theta _d (deg.)$}&
\textrm {Differential cross section (mb/sr)} \\
\hline
   36.4     &    21.9 $\pm$ 0.9    \\
   41.9     &    12.7 $\pm$ 0.7    \\
   46.9     &    9.41 $\pm$ 0.57    \\
   51.6     &    9.17 $\pm$ 0.56    \\
\hline
\hline
\end{tabular}
\centering
\caption{\label{tab17}
Data from the $^{10}$Be(d,d') reaction in inverse kinematics populating the first excited state of $^{10}$Be at E$_{ex}$~=~3.368 MeV at E$_d$~=~15 MeV.}
\begin{tabular}{cc}
\hline
\hline
\textrm{$\theta _d (deg.)$}&
\textrm {Differential cross section (mb/sr)} \\
\hline
   41.0    &     11.8 $\pm$ 0.4    \\
   45.0    &     9.70 $\pm$ 0.35    \\
   48.7    &     8.63 $\pm$ 0.33    \\
   52.2    &     7.58 $\pm$ 0.31    \\
   55.5    &     6.77 $\pm$ 0.29    \\
   58.7    &     6.27 $\pm$ 0.28    \\
\hline
\hline
\end{tabular}
\centering
\caption{\label{tab18}
Data from the $^{10}$Be(d,d') reaction in inverse kinematics populating the first excited state of $^{10}$Be at E$_{ex}$~=~3.368 MeV at E$_d$~=~18~MeV.}
\begin{tabular}{cc}
\hline
\hline
\textrm{$\theta _d (deg.)$}&
\textrm {Differential cross section (mb/sr)} \\
\hline
   36.9    &     8.74 $\pm$ 0.51    \\
   40.4    &     11.0 $\pm$ 0.6    \\
   43.7    &     8.41 $\pm$ 0.50    \\
   46.7    &     6.52 $\pm$ 0.44    \\
   49.6    &     5.85 $\pm$ 0.42    \\
   52.4    &     5.29 $\pm$ 0.39    \\
   55.1    &     6.15 $\pm$ 0.43    \\
\hline
\hline
\end{tabular}
\end{table}
\begin{table}
\centering
\caption{
Data from the $^{10}$Be(d,d') reaction in inverse kinematics populating the first excited state of $^{10}$Be at E$_{ex}$~=~3.368~MeV at E$_d$~=~21.4~MeV.}\label{tab19}
\begin{tabular}{cc}
\hline
\hline
\textrm{$\theta _d (deg.)$}&
\textrm {Differential cross section (mb/sr)} \\
\hline
32.4 &	6.13 $\pm$ 0.22\\
37.2 &	7.92 $\pm$ 0.27\\
40.9 &	8.64 $\pm$ 0.28\\
44.1 &	7.24 $\pm$ 0.27\\
47.1 &	6.09 $\pm$ 0.27\\
50.0 &	6.03 $\pm$ 0.28\\
52.7 &	6.92 $\pm$ 0.27\\
55.4 &	7.48 $\pm$ 0.28\\
57.9 &	5.45 $\pm$ 0.23\\
\hline
\hline
\end{tabular}
\end{table}